\documentclass{aa}  
\usepackage{capt-of}
\usepackage{graphicx}
\usepackage{xcolor}
\def\arcsec{\hbox{$^{\prime\prime}$}}
\def\utw{\smash{\rlap{\lower5pt\hbox{$\sim$}}}}
\def\udtw{\smash{\rlap{\lower6pt\hbox{$\approx$}}}}

\def\farcm{\hbox{$.\mkern-4mu^\prime$}}
\def\farcs{\hbox{$.\!\!^{\prime\prime}$}}

\def\diameter{{\ifmmode\mathchoice
{\ooalign{\hfil\hbox{$\displaystyle/$}\hfil\crcr
{\hbox{$\displaystyle\mathchar"20D$}}}}
{\ooalign{\hfil\hbox{$\textstyle/$}\hfil\crcr
{\hbox{$\textstyle\mathchar"20D$}}}}
{\ooalign{\hfil\hbox{$\scriptstyle/$}\hfil\crcr
{\hbox{$\scriptstyle\mathchar"20D$}}}}
{\ooalign{\hfil\hbox{$\scriptscriptstyle/$}\hfil\crcr
{\hbox{$\scriptscriptstyle\mathchar"20D$}}}}
\else{\ooalign{\hfil/\hfil\crcr\mathhexbox20D}}%
\fi}}

\usepackage{subfig}
\usepackage{txfonts}
\bibpunct{(}{)}{;}{a}{}{,} 
\begin{document}

   \title{The radio relic in Abell 2256: \\
      overall spectrum and implications for electron acceleration}


   \author{M. Trasatti
          \inst{1} \and H. Akamatsu \inst{2} \and L. Lovisari \inst{1} \and U. Klein \inst{1}, 
           A. Bonafede \inst{3} \and M. Br\"uggen \inst{3} \and D. Dallacasa \inst{4,5} \and Tracy Clarke \inst{6}
          }
 \authorrunning{M. Trasatti et al}
          
   \institute{Argelander Institut f\"ur Astronomie, Universit\"at Bonn,
              Auf dem H\"ugel 71, 53121 Bonn, Germany\\
              \email{trasatti@astro.uni-bonn.de}
         \and SRON Netherlands Institute for Space Research, Sorbonnelaan 2, 3584 CA Utrecht, The Netherlands
         \and Hamburger Sternwarte, Universit\"at Hamburg, Gojenbergsweg 112, D-21029 Hamburg, Germany
         \and Dipartimento di Astronomia, Universit\`a di Bologna, via Ranzani 1, I-40127 Bologna, Italy
         \and INAF – Istituto di Radioastronomia, via Gobetti 101, I-40129 Bologna, Italy
         \and Naval Research Laboratory, 4555 Overlook Ave, SW, Washington, DC 20375, USA
             }

   \date{Received XXXX; accepted 01/11/2014}


\abstract
   {Radio relics are extended synchrotron sources thought to be produced by shocks in the outskirts of merging galaxy clusters. The cluster Abell 2256 hosts one of the
   most intriguing examples in this class of sources. It has been found that this radio relic has a rather flat integrated spectrum at low frequencies that would imply
   an injection spectral index for the electrons that is inconsistent with the flattest allowed by the test particle diffusive shock 
   acceleration (DSA). 
   }
   {We aim at testing the origins of the radio relic in Abell 2256.}
   {We performed new high-frequency observations at 2273, 2640, and 4850 MHz. Combining these new observations with images available in the literature, we
    constrain the radio integrated spectrum of the radio relic in Abell 2256 over the widest sampled frequency range collected so far for this class of objects (63 -10450 MHz).
    Moreover, we used X-ray observations of the cluster to check the temperature structure in the regions around the radio relic.}
   {We find that the relic keeps an unusually flat behavior up to high frequencies. Although the relic integrated spectrum between 63 and 10450 MHz is not inconsistent with a single power law with $\alpha_{63}^{10450}= 0.92\pm 0.02$, 
   we find hints of a steepening at frequencies > 1400 MHz. The two frequency ranges 63-1369 MHz and 1369-10450 MHz are, indeed, best represented by two different power laws, with 
      $\alpha_{63}^{1369}= 0.85\pm 0.01$ and $\alpha_{1369}^{10450}= 1.00\pm 0.02$. 
      This broken power law would require special conditions to be explained in terms of test-particle DSA, e.g., non-stationarity of 
      the spectrum, which would make the relic in A2256 a rather young system, and/or non-stationarity of the shock. On the other hand, the single power law would make of this 
      relic the one with the flattest integrated spectrum known so far, even flatter than what allowed in the test-particle approach to DSA.
      We find a rather low temperature ratio of $T_2/T_1 \sim 1.7$ across the G region of the radio relic and no temperature jump across the H region.
      However, in both regions projection effects might have affected the measurements, thereby reducing the contrast. 
    }
   {}


   \maketitle
  
%

\section{Introduction}\label{sec:intro}
A fraction of galaxy clusters exhibit diffuse Mpc-scale 
synchrotron emission (referred to as radio halos and radio relics) 
not related to any particular cluster galaxy \citep[for reviews see][]{2012A&ARv..20...54F, 2012SSRv..166..187B}. 
This emission manifests itself by the presence of relativistic electrons 
($\sim$GeV) and weak magnetic fields ($\sim$ $\mu$G) in the intracluster 
medium (ICM), together with the hot thermal plasma emitting X-rays. 
Radio halos permeate the central Mpc$^3$ of galaxy clusters and 
the radio emission usually follows the roundish X-ray emission from the thermal gas. 
Radio relics are more irregularly shaped and are located at the 
 clusters periphery. 
They are usually further 
subdivided into three classes: radio gischt, radio 
phoenices, and AGN relics \citep[see][]{2004rcfg.proc..335K}, depending 
on their characteristics and proposed origin (as described below). 
The combination of the Mpc size of such sources and the relatively short radiative lifetime of the emitting electrons
implies the need for some form of in situ 
production or (re-)acceleration of the electrons in all these sources, even though the underlying 
physical mechanisms are thought to be different for the different classes of sources.
Moreover, these diffuse radio emitting regions are mostly found in unrelaxed clusters, suggesting 
that cluster mergers play a key role in producing them.\\ 
Radio gischt are large, extended 
arc-like sources, believed to be synchrotron emission from electrons accelerated or re-accelerated 
in merger or accretion shocks through diffusive 
shock acceleration \citep[DSA, Fermi-I process; see][]{1998A&A...332..395E, 2011ApJ...734...18K}. A textbook example of such giant radio relic has been observed 
in the galaxy cluster CIZA J2242.8+5301 \citep{2010Sci...330..347V, 2013A&A...555A.110S}. Radio phoenices are believed to be the result of 
the re-energization via adiabatic compression, triggered by shocks, of fossil plasma
from switched-off AGN radio galaxies \citep{2001ASPC..250..454E, 2002MNRAS.331.1011E}. 
The relativistic plasma of AGN origin had the time to age and without the re-energization would not longer be visible at the currently observable radio frequencies.
An example of a radio phoenix has been found in the galaxy cluster A2443 \citep{2011AJ....141..149C}.
AGN relics are indeed such fossil radio galaxies where the AGN switched off more recently and no re-energization occurred.
The plasma is still emitting at observable radio frequencies, and it simply evolves passively until it becomes invisible in the radio window \citep{1994A&A...285...27K}.
\subsection*{Models and diagnostics}
The proposed formation mechanisms differ in the predictions of the morphological 
and spectral characteristics of the different classes of relics. DSA of both thermal and pre-accelerated electrons, should produce larger and more peripheral structures, 
with strong polarization, and pure power-law integrated synchrotron spectra \citep{2012SSRv..166..187B}. Fermi processes naturally predict an injection 
power-law energy distribution for the accelerated electron population
of the form\footnote{The distribution is truncated at high energy by the existence of a maximum energy to which electrons can be accelerated.} 
$f(E) \propto E^{-\delta_{inj}}$. From synchrotron theory, the emission produced by this population of
electrons is also described by a power law\footnote{The synchrotron spectrum has an exponential cutoff at high frequency, reflecting the truncation in the particle distribution.}, 
\begin{equation} \label{eq:alfa_inj}
 S(\nu) \propto \nu^{-\alpha_{inj}} \qquad \text{with} \qquad \alpha_{inj}=\frac{\delta_{inj}-1}{2}.
\end{equation}
Emitting particles are naturally subject to energy losses. These losses are governed by many physical factors
such as the properties of the magnetic field \citep[see][for a description of the different models of electron aging]{1962SvA.....6..317K, 1973A&A....26..423J, 1994A&A...285...27K}.
The absence of any constant injection of new electrons would lead to a cutoff in the high-energy region of the integrated spectrum, moving toward lower frequencies 
in the course of time. The presence of
constant injection of particles with the same energy spectrum, on the other hand, would eventually mask the cutoff, 
leading instead to a break with a change of 0.5 in the spectral index of the integrated emission \citep[continuous injection model, ][]{1962SvA.....6..317K}:
\begin{equation} \label{eq:alfa_obs}
\alpha_{obs}=\alpha_{inj}+0.5 . 
\end{equation}
This condition translates, in case of DSA, in the assumption that the properties of the shock remain unchanged
(stationarity for the shock).
If the shock has been present in the ICM for a time exceeding the electron cooling time, 
 a single power law with spectral index $\alpha_{obs}$ is expected for the integrated radio spectrum (stationarity for the spectrum).
 The observed spectral indices reported in Table 4 of \citet{2012A&ARv..20...54F}
for straight integrated spectra range from 1.1 to 1.6.
A gradient is expected in the spectral-index distribution across the source, with the flattest values marking the position of the shock 
front where the particle get accelerated, and the steepening showing the radiative losses as the electrons are advected away from the shock.
Such a gradient is clearly observed in the radio relic in CIZA J2242.8+5301 \citep{2010Sci...330..347V, 2013A&A...555A.110S}.\\
In case of revival via adiabatic compression of old radio plasma left behind by 
radio galaxies and pushed towards the cluster outskirts by buoyancy, we expect instead more filamentary and smaller radio structures ($< 50$~kpc), again strongly 
polarized, but with steeper and curved integrated spectra due to the already aged 
population of electrons that are re-accelerated. In fact the adiabatic compression would just 
shift the already aged spectrum at high energies upward, without modifying the spectral slope. Even steeper spectra are expected in case of AGN relics.
The average spectral indices reported in Table 4 of \citet{2012A&ARv..20...54F}
for integrated spectra with measured steepening, range from 1.7 to 2.9.\\
With these ingredients, detailed studies of the integrated spectrum 
and of the spectral-index distribution across the sources, allow us to test 
the current models and study the shock properties in case of DSA. 
This is accomplished by observations made over a broad range of frequencies. 
However, an accurate measurement of the integrated spectra of radio relics 
is a difficult task. These sources usually contain a number of discrete sources, 
whose flux density needs to be carefully subtracted from the total diffuse 
emission. This requires high-resolution imaging at many frequencies using radio interferometers. However,
increasing the observing frequencies, interferometers encounter the technical problem of the
missing short spacings that makes them "blind" to very extended structures. On the other hand, single dishes
are optimal to catch all the emission from a field but they lack angular resolution. Indeed, integrated spectra over a wide range of frequencies are available in the literature
only for few of these objects \citep[see][]{2012A&ARv..20...54F}.\\
An independent measure of the properties of shocks is provided by deep X-ray observations. Through the measurements of temperature and/or 
pressure jumps at the location of the shock, properties such as the shock Mach number $M$ and the shock compression ratio $C$ can be inferred \citep[see review by][]{2012SSRv..166..187B}. 
In the test particles approximation\footnote{When the dynamical feedback of the Cosmic Rays electrons pressure is ignored.} of DSA, 
if the particle diffusion is specified, the shock Mach number is the primary parameter 
that determines the efficiency of the acceleration mechanism and the energy distribution of the particles at injection \citep{2010ApJ...721..886K}. In this case, 
a simple direct relation between the shock Mach number $M$ and the injection index $\delta_{inj}$ of the energy electrons distribution exists:
\begin{equation} \label{eq:delta_inj}
 \delta_{inj}=\frac{2(M^2+1)}{(M^2-1)} .
\end{equation}
However, radio relics are usually observed in the outskirts of clusters where the very low density of electrons ($n_e < 10^{-4} cm^{-3}$) make the detection of
shocks in the X-ray very challenging \citep{2011arXiv1112.3030A}. Indeed, 
a few clear X-ray shock detections are known in the literature \citep[see review by][]{2012SSRv..166..187B}. In conclusion, multifrequency radio measurements, combined with
deep X-ray observations, allow a search for and a proper study of these shocks to test the shock-origin model for relics.
\subsection*{The case of Abell 2256}
One of the most intriguing clusters hosting both a radio relic and a radio halo is the 
galaxy cluster A\,2256 ($z = 0.058$). 
The radio relic emission in this cluster differs in many aspects from the textbook examples of radio gischt in merging clusters, e.g., in 
CIZA J2242.8+5301 \citep{2010Sci...330..347V} and in A3376 \citep{2006Sci...314..791B}. 
The A\,2256 relic emission is, indeed, dominated by a complex filamentary structure as confirmed by new wide-band VLA 
observations published during the reviewing process of the present paper \citep{2014arXiv1408.5931O}. 
It is moreover characterized by an unusually large aspect ratio, being nearly as 
wide as it is long, and by an unusual proximity to the cluster center respect to the majority of giant relics known in the literature.
It also shows all 
typical signatures of a merging cluster system although its dynamical state is
not yet completely understood. 
This cluster has been the first observed with 
LOFAR at very low frequencies (20 - 63~MHz) by \citet{2012A&A...543A..43V}. They collected data up to 1400 MHz
and found a radio integrated spectrum for the relic that can be described by a power law with an unusual flat spectral index $\alpha_{63}^{1369} = 0.81 \pm 0.03$.
The occurrence of similar flat spectral indeces have been reported by \citet{2010ApJ...718..939K} in the frequency range 150-1369 MHz. 
Assuming stationary conditions in the test-particle case of DSA, this would require an injection spectral index 
which is not consistent with the flattest possible injection spectral index from DSA .
Indeed, a direct consequence of the test-particle approach to DSA is that
in the limit of strong shocks ($M  \gg 1$) the particle index $\delta_{inj}$ approaches a asymptotic value of 2. This means that particle
energy distribution produced by test-particle DSA cannot be flatter than 2 (it must be $\delta_{inj} \gtrsim 2$). As a consequence the synchrotron spectra at injection 
cannot be flatter than 0.5 ($\alpha_{inj} \gtrsim 0.5$). So, we should not observe relics with spectra $\alpha_{obs} \lesssim 1$.
The flat spectrum could be reconciled with shock acceleration if the shock has been produced very recently ($\sim 0.1$ Gyr ago) and stationarity has not been reached yet.
In this case a steepening of the integrated spectrum is expected at frequencies $\gtrapprox$ 2000 MHz.\\
In this paper we present new high-frequency radio observations (Sect. \ref{sec:radio_obs}) of A\,2256 at 2273, 2640 and 4850 MHz MHz performed both with an interferometer 
(the Westerbork Synthesis Radio Telescope, WSRT)
and a single dish (the Effelsberg-100m Telescope), complemented by X-ray observations (Sect. \ref{sec:xray_obs}) performed with the Suzaku and XMM-Newton satellites.
In Sect. \ref{sec:radio_results} we present a new determination of the relic radio spectrum over the widest sampled frequency range collected 
so far for this kind of object (63 MHz-10450 MHz)\footnote{A very recent paper reports the first observation of a radio relic at 16 GHz \citep{2014arXiv1403.4255S}.}.
In Sect. \ref{sec:xray_results} we show the ICM temperature in regions across the radio relic emission. In Sect. \ref{Sec:SZ} we consider the effect of the thermal Sunyaev-Zeldovich (SZ) decrement 
on our flux density measurements at high frequencies.
Discussion and conclusions are presented in Sects. \ref{sec:disc} and \ref{sec:conclu}. \\
We adopted the cosmological parameters $H_0=$71 km s$^{-1}$ Mpc$^{-1}$, $\Omega_{\Lambda}$=0.73 and $\Omega_m$=0.27 \citep{2003ApJS..148....1B}, which provide a linear scale of 1.13 kpc arcsec$^{-1}$
at the redshift of A\,2256.
\section{Radio observations and data reduction}\label{sec:radio_obs}
A\,2256 was observed with the WSRT at 2273 MHz and with the Effelsberg 100-m telescope at 2604 and 4850 MHz.\\
In this section we present the observations and the main steps of the calibration and image-making process.
All the observations include full polarization information. In this paper we focus on the total intensity properties of the cluster. 
We postpone a detailed local analysis based on polarization properties and spectral index maps 
to a forthcoming paper (Trasatti et al. in prep.).
\subsection{WSRT observations}\label{subsec:WSRT}
For this project we choose for the WSRT the maxi-short configuration which has optimized imaging performance for very extended sources. 
The receiver covers the frequency range from 
2193 MHz to 2353 MHz
with eight contiguous intermediate frequencies (IFs) of 20 MHz width each; the resulting central frequency is 2273 MHz and the total bandwidth is 160 MHz.
We are potentially sensitive to emissions on scale up to $\sim$13$^\prime$ with a full resolution 
of $\sim$9$^{\prime\prime}$.\\
The main limiting factor of the field of view is 
the effect of the primary beam attenuation. For the WSRT this can be described by the function  
$cos^6(c \cdot \nu \cdot r)$  where r is the distance from the pointing center in degrees, $\nu$ is the observing frequency in GHz and the constant c= 68 is, to first order, 
wavelength independent at GHz frequencies (declining to  c=66 at 325 MHz and c = 63 at 4995 MHz). The resulting field of view at 2273 MHz is 0.37$^\circ$.
In order to image a field big enough to recover the extended emission in A\,2256, the observations were carried out in the mosaic mode. Three different pointing centers were 
chosen (details in Table \ref{table:WSRT_obs}). In order to have a good uv coverage for each pointing, the observations were performed switching the telescope from one pointing 
to another every five minutes, having four hours of observations for each pointing for a total of twelve hours for the entire cluster. 
The observations were carried out on the 25th January 2003.
The excellent phase stability of the system allow us to observe primary calibrators only at the beginning and the end of an observation to calibrate WSRT data. 
3C286 and 3C48 were observed for this purpose.\\
\begin{table*}[ht]
\caption{WSRT Observational parameters.} 
\centering 
\begin{tabular}{c c c c c c} 
\hline
\hline
\noalign{\smallskip}
\multicolumn{2}{c}{Pointing center (J2000)} & Frequency & Bandwidth & Exposure time & Telescope configuration\\
RA & DEC & (MHz) & (MHz) & (h) & \\ 
\noalign{\smallskip}
\hline 
\noalign{\smallskip}
17 01 07.998 & +78 45 03.701 & 2273 & 160 & 4 & maxi-short\\
17 04 25.000 & +78 45 03.701 & 2273 & 160 & 4 & maxi-short\\
17 02 42.300 & +78 34 59.988 & 2273 & 160 & 4 & maxi-short\\
\noalign{\smallskip}
\hline
\end{tabular}
\label{table:WSRT_obs} 
\end{table*}
Flagging, calibration, imaging and self-calibration were performed with the AIPS (Astronomical Image Processing System) package, with standard procedures following the 
guideline provided on the ASTRON web-page\footnote{\url{http://www.astron.nl/radio-observatory/astronomers/analysis-wsrt-data/analysis-wsrt-dzb-data-classic-aips/analysis-wsrt-d .}}.
All the antennas were successful, with some occasional RFI, flagged out in the early stages of data calibration. 
3C286 was used as the main flux density calibrator using the \citet{1977A&A....61...99B} scale (task SETJY in AIPS), which provides flux densities ranging from 
11.74 Jy in the first IF to 11.39 Jy in the eighth IF.
The three pointings were imaged and self-calibrated separately. 
For each pointing we performed three phase-only cycles of self-calibration, followed by a final amplitude and phase self-calibration cycle. The diffuse
emission flux was included in the model for the self-calibration.
A multiresolution clean was performed within the IMAGR task in AIPS to better reconstruct the complex
diffuse emission present in the cluster in the final images of the pointings. 
Images of the Stokes parameter I, U and Q were obtained for each pointing and were then combined together (separately for I, U and Q)
and corrected for the primary beam attenuation with the FLATN task in AIPS providing a central region with a uniform $\sigma$ noise distribution of $\sim$ 0.027 mJy.
The primary beam correction determines an increase of the noise in the outer regions.
\subsection{EFFELSBERG observations}\label{subsec:eff}
Part of the observations were performed with the Effelsberg 100m Telescope.
We used the 11 cm (=2640 MHz) and 6 cm (=4850 MHz) receivers. Single-dish observations do not suffer from the zero-spacing problem, and can trace large scale 
features, although with modest resolution.\\ 
The data reduction of Effelsberg data was performed with the NOD2 software package, following the standard procedures provided on the 
MPIfR web-page\footnote{\url{https://eff100mwiki.mpifr-bonn.mpg.de/doku.php?id=information_for_astronomers:user_guide:reduc_maps .}}.
The raw images of both A\,2256 and the calibrators were partly processed using dedicated pipelines available for each receiver. 
The default strategy to calibrate Effelsberg data is to observe primary calibrators during the session and then use automatic 2-D Gauss fit pipelines
to calculate the factor to scale the final image converting it from mapunit/beam to Jy/beam (task RESCALE in AIPS).
\subsubsection{Observations at 2640 MHz}\label{subsec:13cm}
The Effelsberg 11 cm receiver is a single-horn system equipped with a polarimeter with eight small-band frequency channels, each 10 MHz wide, covering the frequency range 2600-2680 MHz,
plus one broad-band channel, 80 MHz wide, over the same frequency range. The resulting central frequency is 2640 MHz and the total bandwidth is 80 MHz.
The resolution of the observation is 4$\farcm 5\ \times 4\farcm 5\ $.\\
To map the A\,2256 field we used the mapping mode, which consist in rastering the field of interest by moving the telescope, e.g., along longitude (l), 
back and forth, each subscan shifted in latitude (b) with respect to the other. 
At centimeter wavelengths atmospheric effects (e.g., passing clouds) introduce additional emission/absorption while scanning, leaving a stripy pattern along 
the scanning direction (the so-called scanning effects). Rastering the same field along two perpendicular directions (both along longitude and latitude) 
helps in efficiently suppressing these patterns, leading to a sensitive image of the region \citep{1988A&A...190..353E}.
This technique, called basket-weaving technique, helps also in setting the zero-base level. The details of the observations are summarized in Table \ref{table:EFF_obs}.
For each coverage of the field, the receiver provides four images (R, L, U, Q) for each of the nine channels. As circular polarization is generally very weak, the images in R and L 
are very similar and can be averaged in the later steps of data reduction providing the total intensity image.\\
We performed a total of 14 coverages in the longitude direction and 15
coverages in the latitude direction; due to RFI (Radio Frequency Interference) and pointing problems we had to discard a small portion of the data.
The time required to complete one coverage is $\sim$ 17 minutes in both direction, 
so we have a total observing time on source of 8.2 hours. The observations were carried out in the night between the 15th and 16 August 2012.\\
\begin{table*}[ht]
\caption{Effelsberg observational parameters.} 
\centering 
\begin{tabular}{c c c c c} 
\hline
\hline
\noalign{\smallskip}
\multicolumn{2}{c}{Map center (J2000) }         & Frequency  & Bandwidth  &Map size                      \\
 RA                 & DEC                       & (MHz)      &  (MHz)     & ($^\prime \times$ $^\prime$)\\ 
 \noalign{\smallskip}
\hline 
\noalign{\smallskip}
 17 04 00        &+78 04 00  & 2640             & 80         &   48$\times$48  \\
 17 04 00        &+78 04 00  & 4850             & 500        &   40$\times$35 \\
\noalign{\smallskip}
\hline
\end{tabular}
\label{table:EFF_obs} 
\end{table*}
3C286 and 3C48 were used as absolute flux density calibrators using the \citet{1977A&A....61...99B} scale
 that provide flux densities at 2640 MHz of 10.65 Jy and 9.38 Jy respectively for the two calibrators.
\subsubsection{Observations at 4850 MHz}\label{subsec:6cm}
The Effelsberg 6 cm receiver is a double-horn system, with the two feeds fixed in the secondary focus with a separation of 6 cm, each with 
one broad-band (500 MHz) frequency channel in the range 4600-5100 MHz.
The resulting central frequency is 4850 MHz and the total bandwidth is 500 MHz. The resolution of the observation is 2$\farcm43\ \times 2\farcm43$.\\
Multihorn systems use a different technique to overcome the scanning effect problem.
The scanning is done in an azimuth-elevation coordinate system, and must be done only in azimuth direction so that all horns will cover the same sky area 
subsequently. At any instant each feed receives the emission from a different part of the sky but they are affected by the same atmospheric effects, 
which then cancel out taking the difference signal between the two feeds (Emerson et al. 1979).
Similarly to the 11 cm receiver, data in (R, L, U, Q) are provided for each of the two horn.\\
We performed a total of 25 coverages of the A\,2256 field, 15 during the night between the 22nd and the 23rd of June and 10 on the 26th of June 2011.
Due to RFI problems only 22 coverages could be used.\\
For the calibration we observed 3C286 and NGC7027 during the session. 
The flux densities used for the two calibrators are 7.44 Jy \citep[from][]{1977A&A....61...99B} and 5.48 Jy \citep[from][]{2000A&AS..145....1P} respectively.\\
\section{X-ray observations and data reduction}\label{sec:xray_obs}
In this section we present the X-ray observations of A\,2256 performed with Suzaku and XMM-Newton and the main steps of data reduction.\\
Suzaku observed the radio relic region in A\,2256 \citep[OBSID: 801061010, ][]{2011PASJ...63S1009T}
with an exposure time of 95.2 ks. The satellite
X-ray Imaging Spectrometer (XIS: Koyama et al. 2007) has a very low
detector background, which allows us to investigate weak X-ray
emission targets such as cluster outskirts \citep[see][for a review]{2013SSRv..177..195R}.
The XIS was operated in the normal
clocking, 3$\times$3 and 5$\times$5 mode. 
To increase the signal-to-noise ratio we filtered the dataset using a geomagnetic cosmic-ray cut-off rigidity (COR) $> $ 8 GV. The filtered exposure time is 89.2 ks.
The data were processed using standard Suzaku pipelines \citep[see][for more details]{2012PASJ...64...67A}.\\
We complemented Suzaku observations with XMM-Newton observations retrieved from the archive (OBSID: 0141380101 and OBSID:0141380201) 
and reprocessed with SAS v11.0.1. 
The data were heavly affected by soft proton flares. The data were
cleaned for periods of high background due to soft proton flares with a two stage filtering process 
\citep[see][for more details on the cleaning process]{2009A&A...508..191L, 2011A&A...528A..60L}. In this screening process bad pixels have been excluded and only 
event patterns 0-12 for the MOS detectors and 0 for the pn detector were considered. The filtered exposure time is $\sim$19 ks for MOS1, $\sim $20 ks for MOS2 and 
$\sim$9 ks for pn. \\
For both satellites the background emission can be described as the sum of a particle background component and a sky background component.
The former is produced by the interaction of high-energy particles with the detectors. The latter can be subdivided into at least two thermal components, 
one unabsorbed due to the Local Hot Bubble (LHB: kT$\sim$0.08 keV) and one absorbed due to the Milky Way Halo (MWH: kT$\sim$0.3 keV), 
and a power-law component due to the Cosmic X-ray Background (CXB: $\Gamma$=1.41). \\
The particle background has been modeled and subtracted from the data of both satellites before the spectral fits presented in Sect. \ref{subsec:Treg}. 
For the Suzaku observations its contribution has been estimated from the Night-Earth database with the ${\it xisnxbgen}$ FTOOLS \citep{2008PASJ...60S..11T}. 
For XMM-Newton the particle component spectra have been extracted from the filter wheel closed (FWC) observations and renormalized by using the out-of-field-of-view events. 
These spectra were supplied as background spectra to the XSPEC fitting routine. \\
Unlike the particle background, the sky background was not subtracted from the data but its different components were modeled together with the ICM emission during the spectral fitting.
To fix the model parameters for the different components in Suzaku observations, we used spectra extracted from a 1 degree offset observation performed with the satellite 
(PI: Kawaharada, OBSID: 807025010).
For XMM-Newton data, we followed the method presented in Snowden et al. (2008) in which the different components are estimated using a 
spectrum extracted from ROSAT data in an annulus beyond the virial radius of the cluster.
The offset spectra were fitted with a sky background model considering the LHB, MWH and CXB components.
In the fitting, we fixed the temperature of the LHB component to 0.08 keV. Abundance \citep{1989GeCoA..53..197A} and redshift of LHB and MWH were fixed to 1 and 0, respectively. 
The temperature of the MWH determined in the fit is 0.21 $\pm$ 0.03 keV.
We also checked for the possibility of an additional "hot foreground" component with
$kT\sim0.6-0.8$ keV (Simionescu et al. 2010) adding another thermal component to the background model described above. 
However, the intensity of this additional component resulted not significant in the offset field and was not included in the background modeling. 
\begin{figure*}[!ht]
  \centering
      \includegraphics[trim=0cm 0cm 0cm 0cm, clip=true, width=17cm]{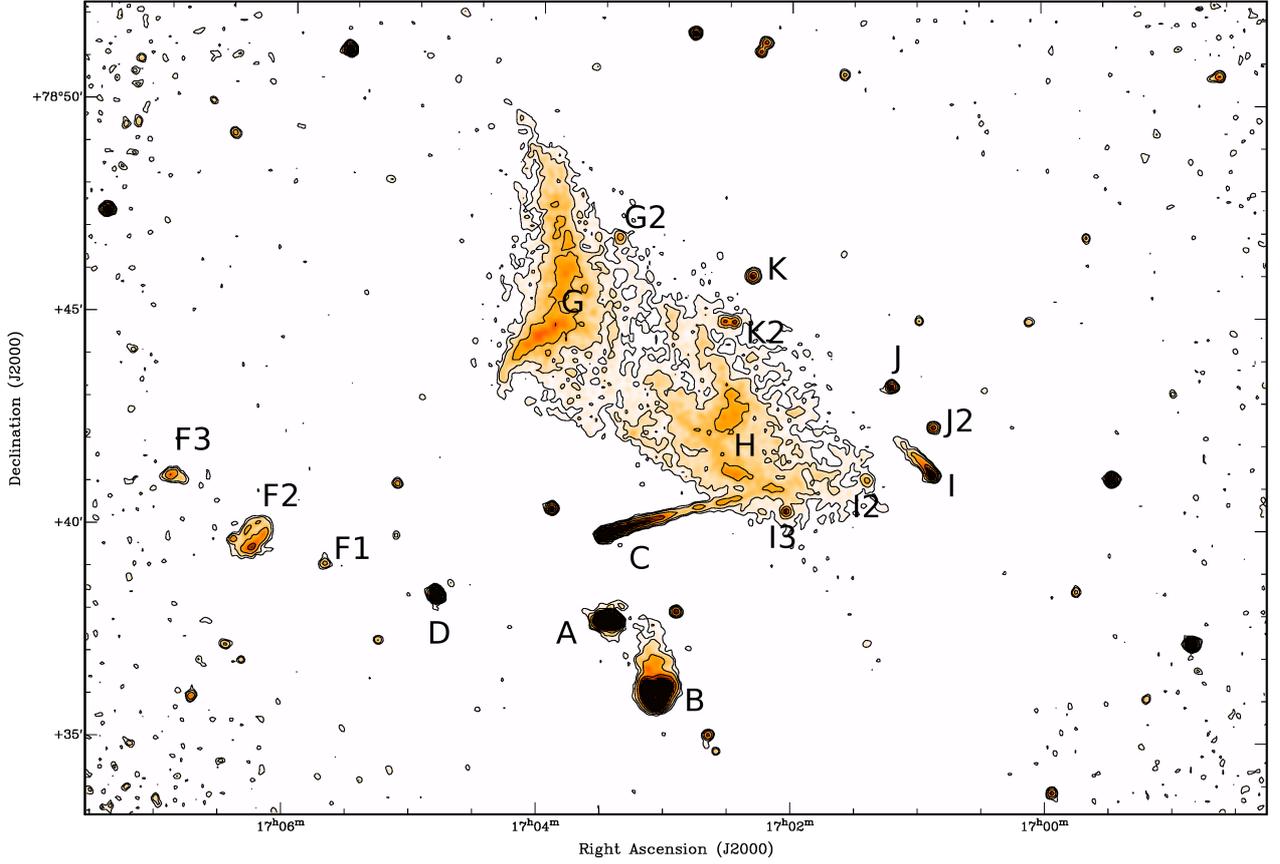}
      \caption{WSRT 2273 MHz total intensity radio image. Contours are drawn at [1, 2, 4, 8] $\times$ 3$\sigma$, with $3\sigma = 8 \cdot 10^{-5}$ Jy/beam and the color scale starts 
      at the same level. The beam size is $9\farcs84 \times 9\farcs44$. The image is corrected for the primary beam attenuation.  }
         \label{fig:WSRT_TI}
   \end{figure*}

\section{Radio analysis and results}\label{sec:radio_results}
  \subsection{Radio images}\label{subsec:radio_images}
 \subsubsection{WSRT image}
In Figure \ref{fig:WSRT_TI} we present the 2273 MHz total intensity WSRT image of the central region of A\,2256. 
The image has been produced with natural weighting of the visibilities in the range [$u_{min}-u_{max}$]=[260 - 21035 $\lambda$]. The shortest spatial frequency sampled 
$u_{min}$ determines the largest spatial scale recovered by this observations $LSS \simeq 1/u_{min}=13\farcm22\ $. 
The image has been corrected for the primary beam attenuation that determines an increase of the noise in the edge of the image.
The high resolution ($9\farcs84\ \times 9\farcs44\ $) allowed us to analyze the substructures of the diffuse relic emission in detail.
The map shows several of the well-known radio features 
 present in the cluster \citep[notation from][]{1979A&A....80..201B, 1994ApJ...436..654R}: the radio relic emission (sources G and H), 
the head-tail sources A, B, C and I, the complex source F (here resolved in the three components F1, F2 and F3), as well as many other discrete sources, some of which
labeled in this paper as I2, I3, K2, G2, J2. 
The radio halo emission present 
in the center of the cluster around source D \citep{2006AJ....131.2900C} is completely filtered out due to a combination of effects: its low 
surface brightness at this frequency, combined with the lack of sampled short spacings in the WSRT observations, that determines the loss of the very extended weak emission.\\
The relic emission exhibits two regions of enhanced surface brightness: a well-defined arc-like region (G region) in the northern part, and a 
less defined region (H region) in the western part (see Fig. \ref{fig:WSRT_TI}). The two regions are connected by a bridge of lower brightness emission.
At full resolution  the entire relic emission covers an area of about $10\farcm6\ \times 5^\prime $. The size reported by 
\citet{2006AJ....131.2900C} at 1.4 GHz and at a resolution of $52^{\prime \prime} \times 45^{\prime \prime}$ is $16\farcm9\ \times 7\farcm8\ $. 
Convolving our WSRT image to the same resolution 
we obtain a similar angular size (not shown). Nevertheless, as we might be anyway missing some of the flux on the most extended scales,
we do not use this image for the computation of the relic integrated spectrum.  
\subsubsection{EFFELSBERG images}
\begin{figure*}[ht]
   \centering
  \includegraphics[trim=0cm 0cm 1.3cm 0cm, clip=true, width=9.1cm]{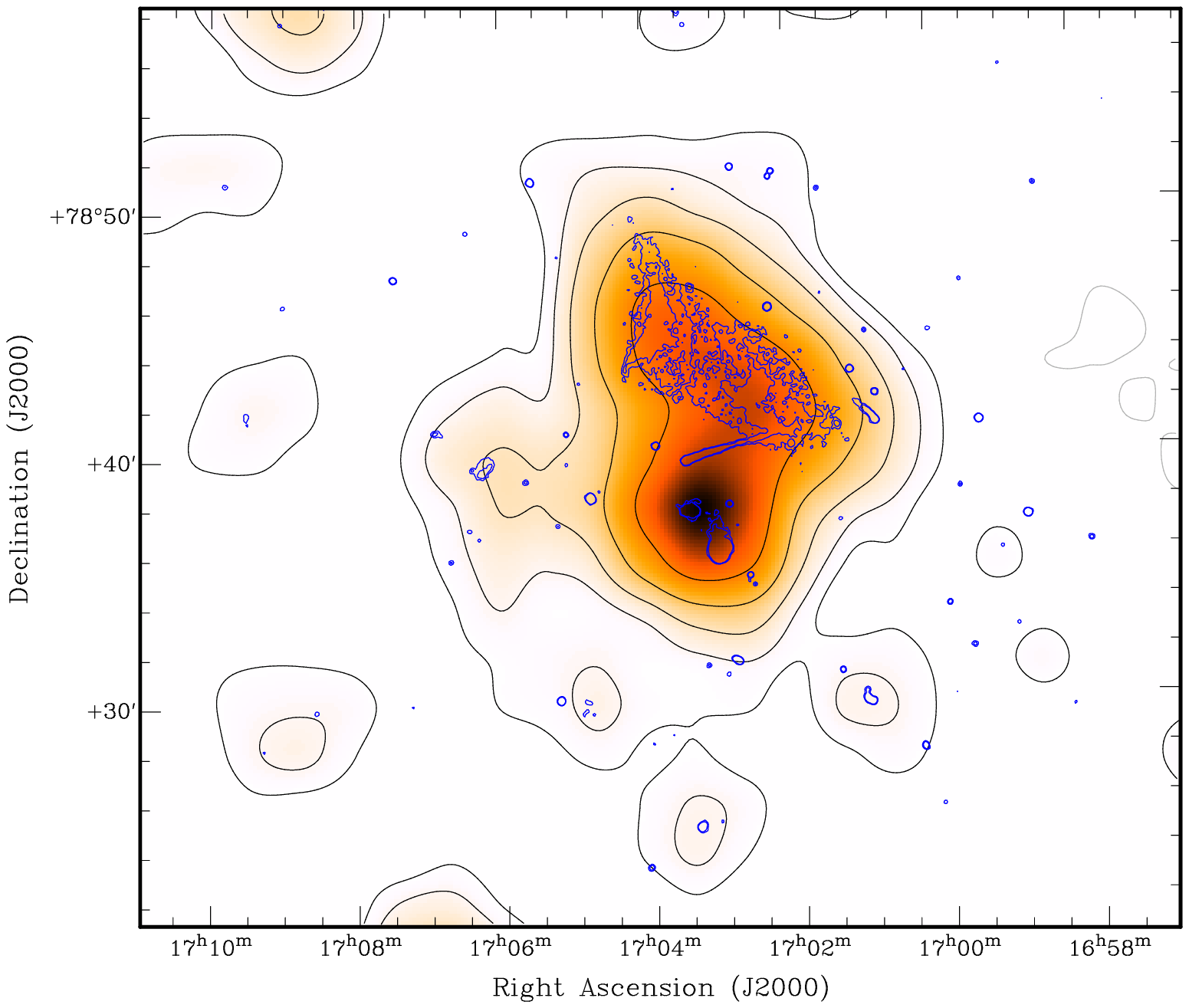}
  \includegraphics[trim=0cm 0cm 1.3cm 0cm, clip=true, width=9.1cm]{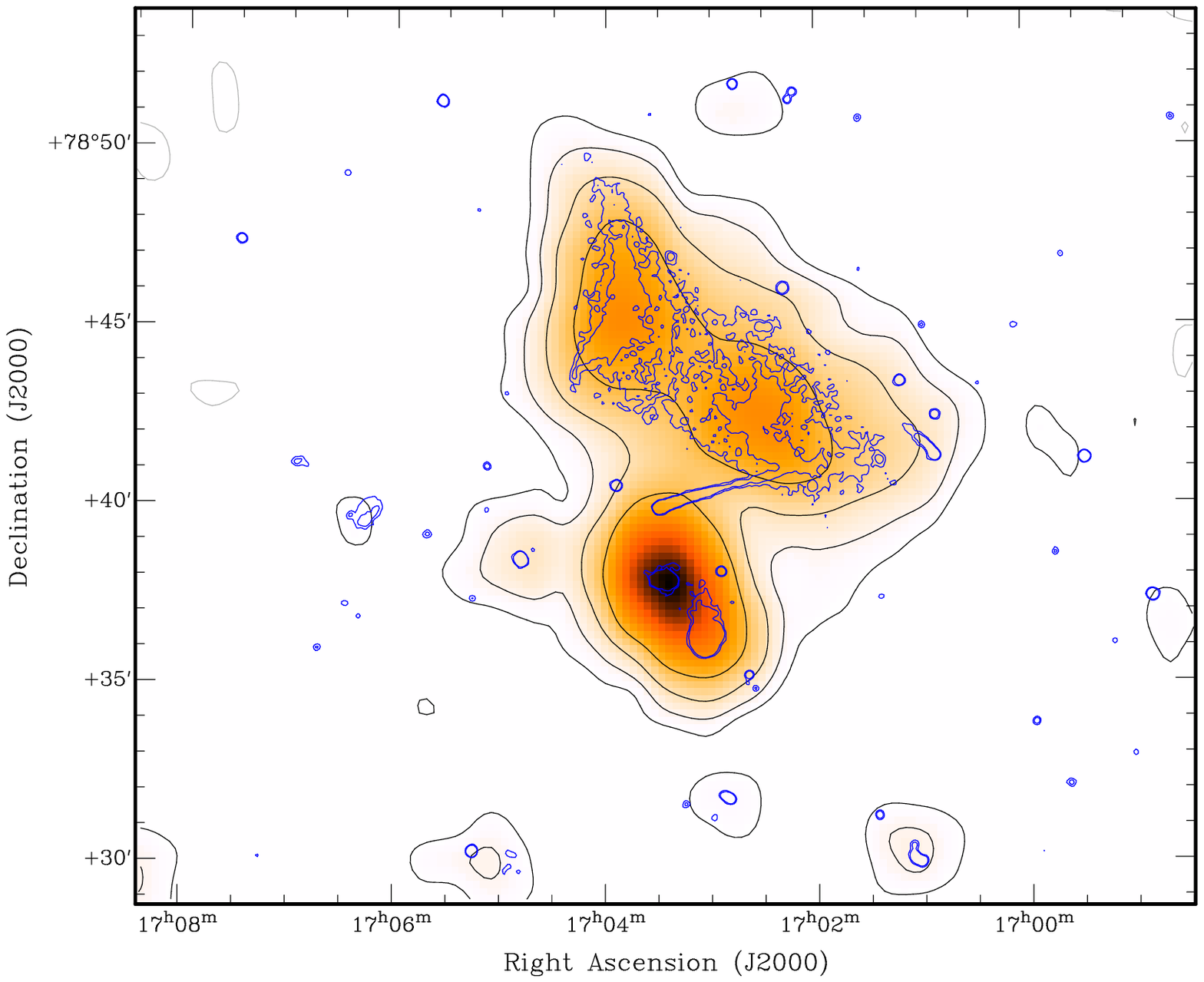}
        \caption{Effelsberg total intensity radio images. {\itshape Left panel}: Effelsberg 2640 MHz total intensity radio image in color scale and black contours 
        drawn at [-1, 1, 2, 4, 8, 16] $\times$ 3$\sigma$, with $3\sigma = 4 \cdot 10^{-3}$ Jy/beam. The beam size is $ 4\farcm4\ \times 4\farcm4\ $. 
        {\itshape Right panel}: Effelsberg 4850 MHz total intensity radio image in color scale and black contours 
        drawn at [-1, 1, 2, 4, 8] $\times$ 3$\sigma$, with $3\sigma = 1.86 \cdot 10^{-3}$ Jy/beam. The beam size is $2\farcm43\ \times 2\farcm43\ $. In both panels
        the blue contours are from the WSRT 2273 MHz radio image not corrected for the primary beam attenuation and are drawn at [1, 2] $\times$ 8$\cdot 10^{-5}$ Jy/beam.}
\label{fig:EFF_TI}
   \end{figure*}

In Figure \ref{fig:EFF_TI} we show the 2640 MHz ({\itshape Left panel}) and 4850 MHz ({\itshape Right panel}) Effelsberg images of A\,2256. In both images, the diffuse emission
 from the relic is mixed up with the emission from the more compact sources present in the field due to the low resolution of the observations. 
 In the 2640 MHz image the relic emission is blended with the emission from the A+B+C complex and from source F. 
 At 4850 MHz it is easier to separate the relic emission from the emission of the A+B complex, but still part of the tail of source C is inevitably superimposed 
 on the Western part 
 of the relic. Moreover, \citet{2008A&A...489...69B} derived a spectrum $\alpha$=1.5$\pm$0.2 for the radio halo between 351 and 1369 MHz. If there is no steepening in the spectrum
 of the halo, we expect a flux density of $\sim$ 37 mJy at 2640 MHz and a flux density of $\sim$ 15 mJy at 4850 MHz. 
 Since a single dish is sensitive to all the emission in the 
 field, its emission is smoothed with the other sources in our Effelsberg images if the halo spectrum keeps straight at these frequencies. 
 Constraining the halo spectrum at high frequencies would require a dedicated careful analysis that is beyond the aims of this paper.
\subsection{Spectral analysis}\label{subsec:spectra} 
For the spectral analysis of the radio relic in A\,2256 we combined our high-frequency observations with images obtained at other frequencies provided by the authors: 
the 351 MHz WSRT image \citep{2008A&A...489...69B}, the 1369 MHz VLA C and D configuration images\footnote{We used the high-resolution 
C configuration data to measure the flux density of discrete sources 
while for the determination of the relic flux density we used the D configuration image.} 
\citep{2006AJ....131.2900C} and the Effelsberg 10450 MHz image \citep{Thierbach...M}.
We moreover got information on the flux densities in the LOFAR image at 63 MHz published by \citet{2012A&A...543A..43V} by the author. 
All the different images were calibrated according to the flux scale of \citet{1977A&A....61...99B} or to its extension to lower frequencies 
\citep[below 408 MHz;][]{PT....1999} \footnote{The overall flux scale for the LOFAR observations at 63 MHz was obtained comparing the 
measured integrated flux densities of five bright sources in the field of view with predicted fluxes partly based on flux densities measurements from the 1.4 GHz NVSS 
\citep{1998AJ....115.1693C} and the 74 MHz VLSS \citep{2007AJ....134.1245C}, both based on the \citet{1977A&A....61...99B} scale.}.
In this way we were able to cover the widest range of frequencies (63 MHz- 10.45 GHz)
used so far for the determination of the spectrum of a radio relic.\\
We estimated the uncertainties $\sigma_S$ on the flux density measurements $S$ with the following formula
 \begin{equation}
  \sigma_S=\sqrt{{\sigma_{rms}}^2 + {\sigma_{cal}}^2} ,
 \end{equation}
where\\
 \begin{itemize}
  \item $\sigma_{rms}=\sigma \times \sqrt{N_{beam}}$ is the error due to the image noise, with $\sigma$ being the image noise level 
  (quoted in the image's captions)
  and $N_{beam}$ the number of beams covered by the source;
  \item $\sigma_{cal}=E_{cal} \times S$ is the error due to calibration uncertainties, determined in turn by two factors: the accuracy of the absolute 
  flux density scale adopted ($\epsilon_{scale}$) and the uncertainties 
  related to the application of such scale to our data (the calibration method, $\epsilon_{cal}$); being these two factors uncorrelated we used 
  $E_{cal}=\sqrt{\epsilon_{scale}^2 +\epsilon_{cal}}^2$.
 \end{itemize}
The spectral data provided by \citet{1977A&A....61...99B} for the flux density calibrators have an absolute uncertainty of 5\%.
For the calibration of the Effelsberg images we performed 4 coverages of 3C286 and 3 coverages of 3C48 at 2640 MHz and 4 coverages of 3C286 and 
3 coverages of NGC7027 at 4850 MHz. A Gaussian fit of the image of a calibrator provides its flux density in map unit. A comparison of this value with the known
calibrator flux density allows us to calculate the factor to translate map unit in physical unit. The slightly different factors deriving from the different coverages of the calibrators were finally 
averaged, separately at the two frequencies.
The standard deviation of these values (at the two frequencies) was used as the term $\epsilon_{cal}$. For the WSRT data we used, instead, 
the dispersion of antenna gains.
 For the term $St_{scale}^2$ we used the standard deviation on the calibration factor derived from the Gaussian 
 fit of the calibrators images produced during the session for the Effelsberg observations, and the dispersion of antenna gains for the WSRT data. 
 This translates into $\epsilon_{cal}$ values of 1.6\% for the Effelsberg 11cm data, of 1.2\% for the Effelsberg 6 cm data and of 2\% for the WSRT 13 cm data. 
 Combining this with the uncertainties on the \citet{1977A&A....61...99B} scale, we end up with $E_{cal}(11cm)$=5.3\%, $E_{cal}(6cm)$=5.2\% and $E_{cal}(13cm)$=5.4\%. 
 For the other images from the literature we assumed a similar value $\epsilon_{cal}$=6\%.
 It should be noted that the self-calibration process might affect the measured flux densities on interferometric images. The uncertainties are therefore possibly 
 larger than estimated here.\\
 Where not otherwise specified, we estimated the uncertainties $\sigma_f$ on the quantities $f$ 
 calculated from measured quantities 
 (e.g., spectral indices and extrapolated flux densities of discrete sources) applying the 
 standard error propagation formula.
The spectra of the total cluster, relic+sources, relic, G and H regions presented in the next sections 
have been determined calculating the flux densities at different frequencies on images convolved to the same lowest resolution available 
($4\farcm4\ \times 4\farcm4\ $) and with the same regions for the integration. The errors associated to the spectral indices are the errors from the fits of the data 
taking into account the uncertainties in the flux density measurements.\\
All the spectra, including those of the discrete sources embedded in the relic emission, are plotted over the same fixed x-axis range 
(frequency range 40 - 14000 MHz) for easy comparison.
\subsubsection {Total cluster emission}\label{subsec:totcl}
We first considered the integrated radio flux density of the entire cluster (halo, relic and discrete sources combined) measuring the flux density in the circular region 
centered at J2000 position $\alpha$ = 17 03 45 $\delta$ = +78 43 00 with 10$^\prime$ radius, as described by \citet{2008A&A...489...69B}.
The cluster radius was determined by \citet{2008A&A...489...69B} as the one at which the derivative of the integrated flux within the circle respect to the radius of the circle
 settles to a constant value.
The measured flux densities at 351, 1369 (D configuration), 2640, 4850 and 10450 MHz are summarized in Table \ref{table:totflux}.
The total cluster radio emission between 351 and 10450 MHz can be modeled by a single power law with spectral index $\alpha$= 1.01 $\pm$ 0.02 
(Fig. \ref{fig:totclustsp}).\\
\citet{2008A&A...489...69B} modeled the cluster flux density as the sum of two spectral components, one due to the halo and the other due to the relic and discrete sources combined. He showed 
that the second term becomes dominant at frequencies above 100 MHz.
Being the radio relic the dominant contributor to the flux density at high frequency, its spectrum at the same frequencies cannot be 
flatter than the total cluster emission spectrum. This 
shows qualitatively that at high frequencies the relic spectrum does not keep the 0.85 slope observed at low frequency by \citet{2012A&A...543A..43V}.
In Sect. \ref{subsec:relic_spec} we will quantify this steepening.
 \subsubsection{Discrete sources }
     \begin{figure}[ht]
   \centering
   \includegraphics[width=8cm]{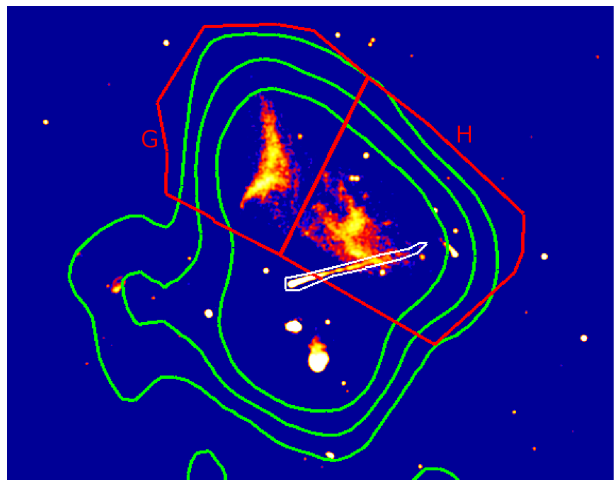}
     \caption{Regions used for the spectra computation. 
     The red region marks the region considered for the relic spectrum computation. The total region is further subdivided into regions G and H. 
     In color scale the WSRT high-resolution image with the green contours from the 2640 MHz Effelsberg image overplotted. The white region mark
     the C source.}
         \label{fig:spec_areas}
   \end{figure}
   \begin{figure}[!h]
   \centering
   \includegraphics[trim=0cm 3cm 0cm 3cm, clip=true, width=10.5cm]{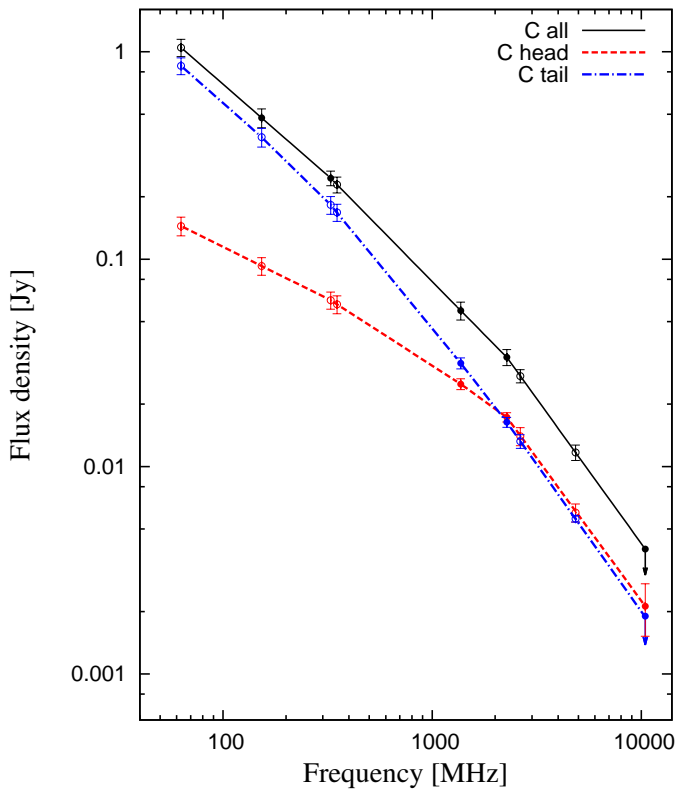}
     \caption{Integrated radio spectrum of source C. Filled circles are measured flux densities while open circles are extrapolated flux densities. The black solid line is the spectrum of the entire
     source; the red dashed line is the source's head spectrum; the blue dotted-dashed line is the spectrum of the tail. See text for more details. }
     \label{fig:Cspe}
   \end{figure}

  \begin{figure}[!h]
   \centering
   \includegraphics[trim=0cm 7cm 0cm 7cm, clip=true, width=10.5cm]{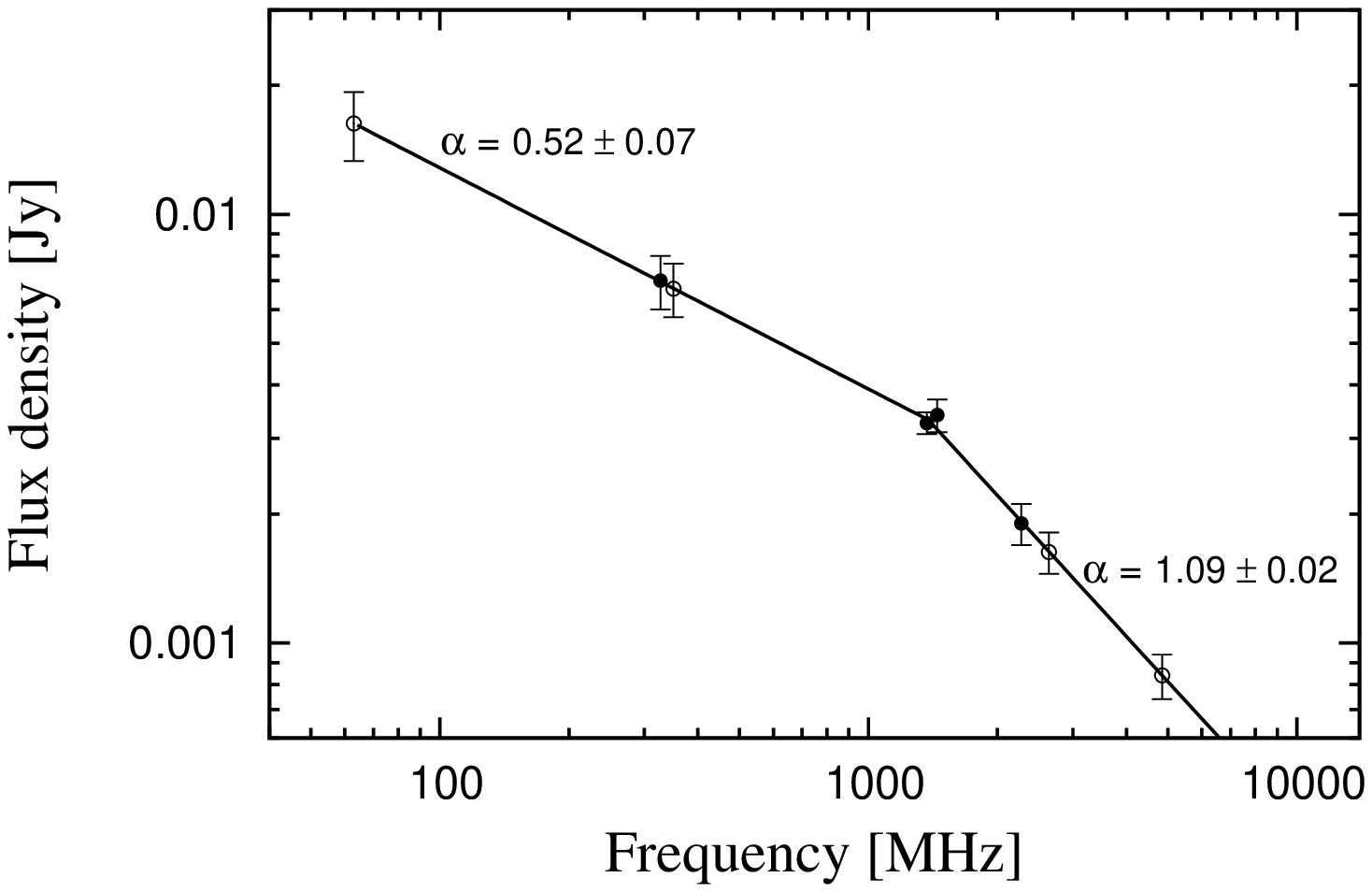}
     \caption{Integrated radio spectrum of source K. Filled circles are measured flux densities while open circles are extrapolated flux densities. }
         \label{fig:Kspe}
   \end{figure}  
 
 \begin{figure}[!h]
   \centering
   \includegraphics[trim=0cm 8cm 0cm 8cm, clip=true, width=10.5cm]{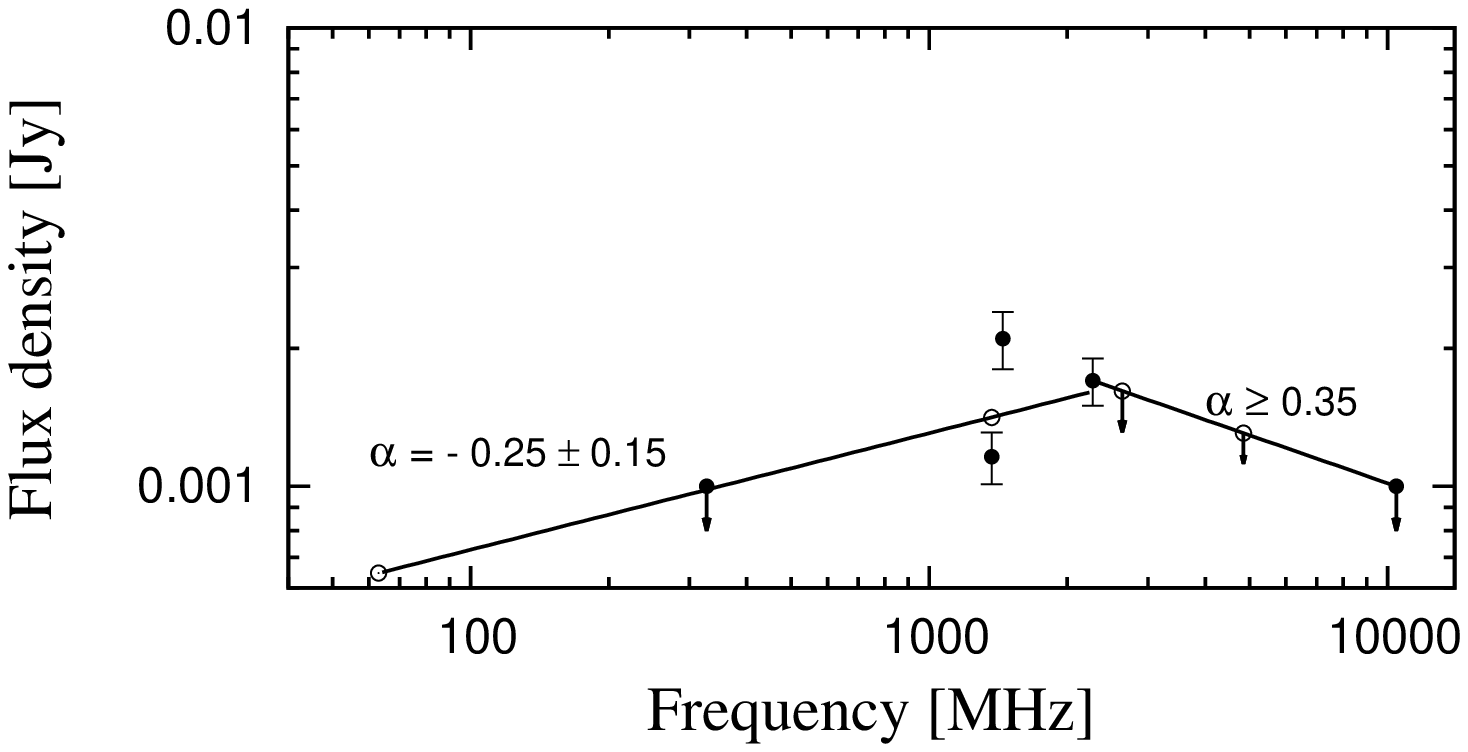}
     \caption{Integrated radio spectrum of source J. Filled circles are measured flux densities while open circles are extrapolated flux densities. }
         \label{fig:Jspe}
   \end{figure}
   
  \begin{figure}[!h]
   \centering
   \includegraphics[trim=0cm 6cm 0cm 6cm, clip=true, width=10.5cm]{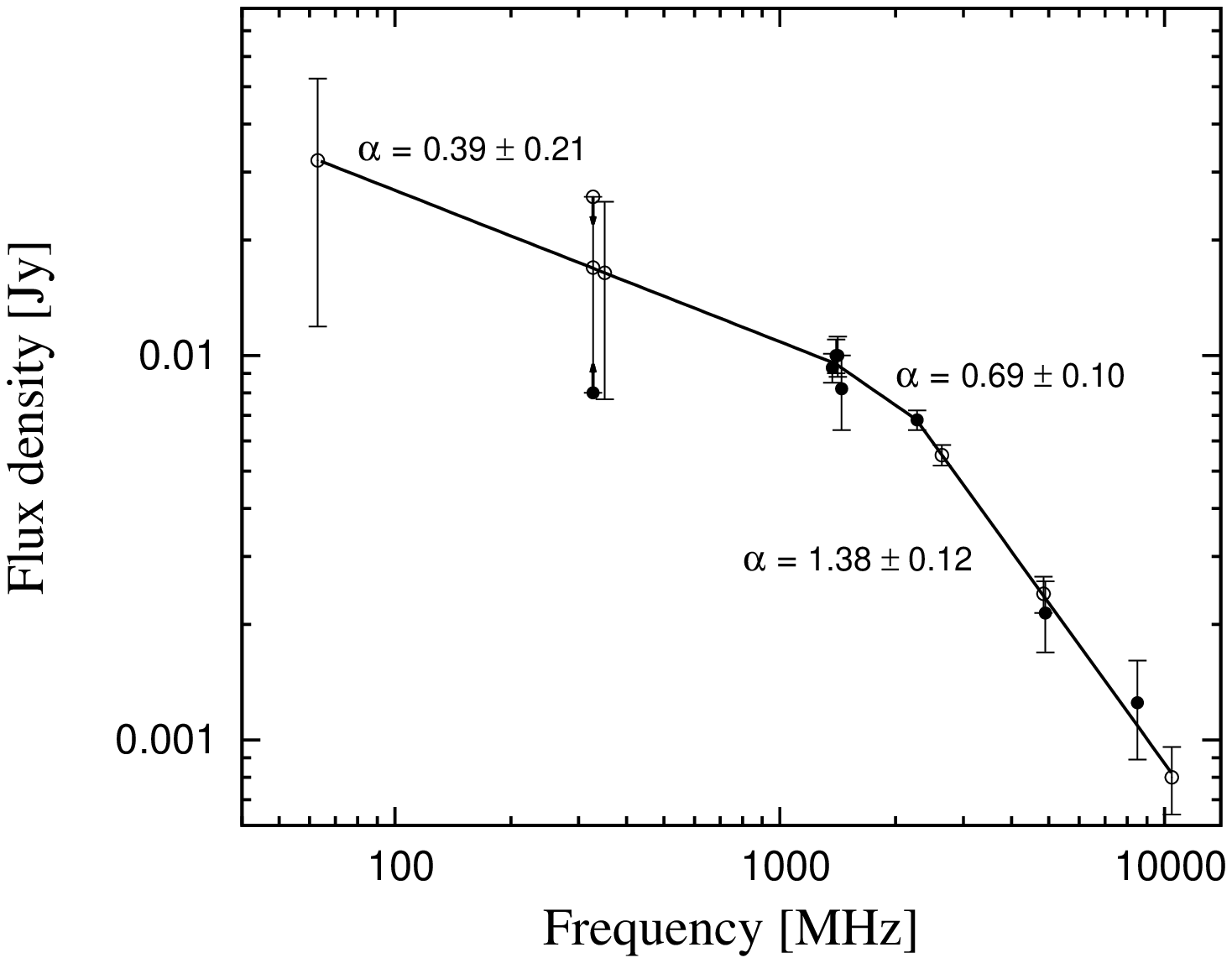}
     \caption{Integrated radio spectrum of source I. Filled circles are measured flux densities while open circles are extrapolated flux densities. }
         \label{fig:Ispe}
   \end{figure}
 
 \begin{figure}[!ht]
   \centering
   \includegraphics[trim=0cm 8cm 0cm 8cm, clip=true, width=10.5cm]{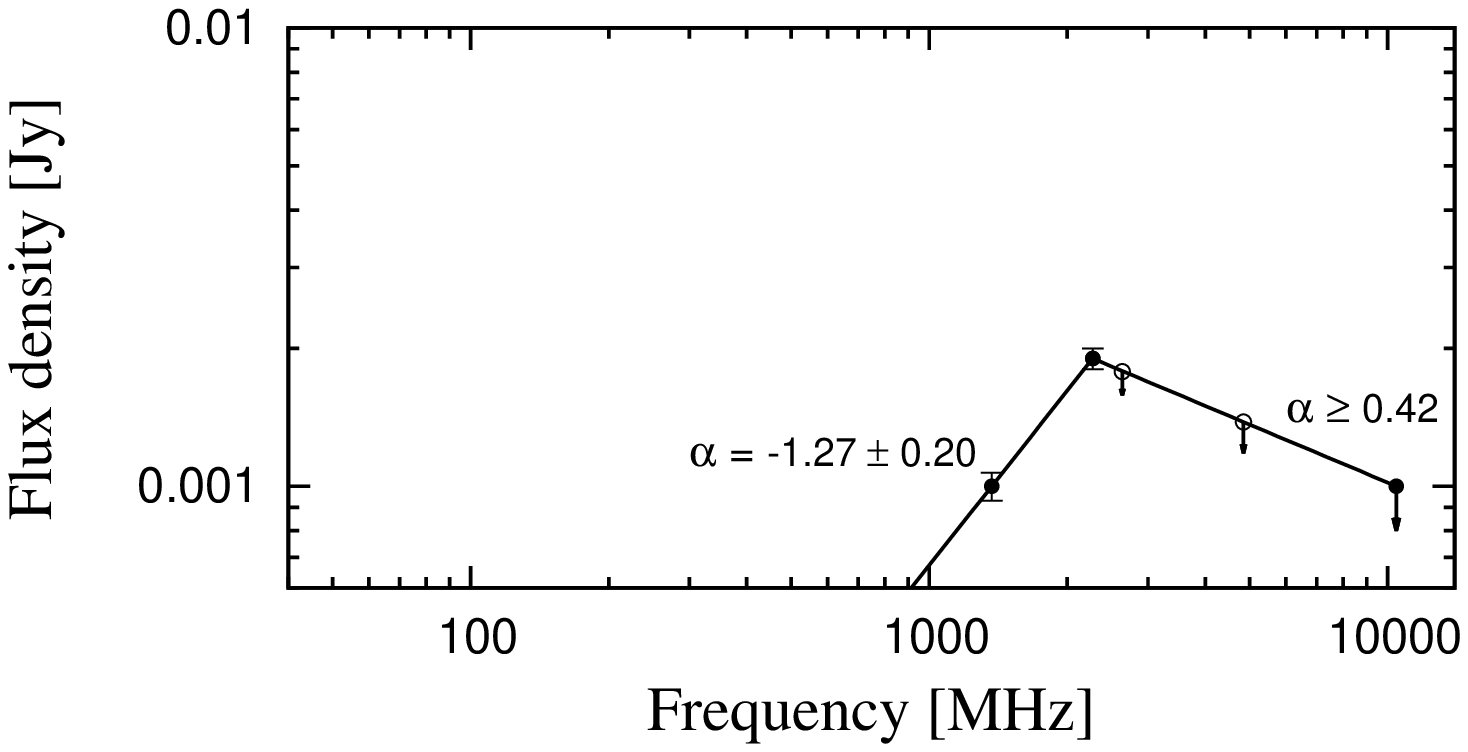}
     \caption{Integrated radio spectrum of source I2. Filled circles are measured flux densities while open circles are extrapolated flux densities.}
         \label{fig:I2spe}
   \end{figure}
 \begin{figure}[!h]
   \centering
   \includegraphics[trim=0cm 1cm 0cm 1.5cm, clip=true, width=10.5cm]{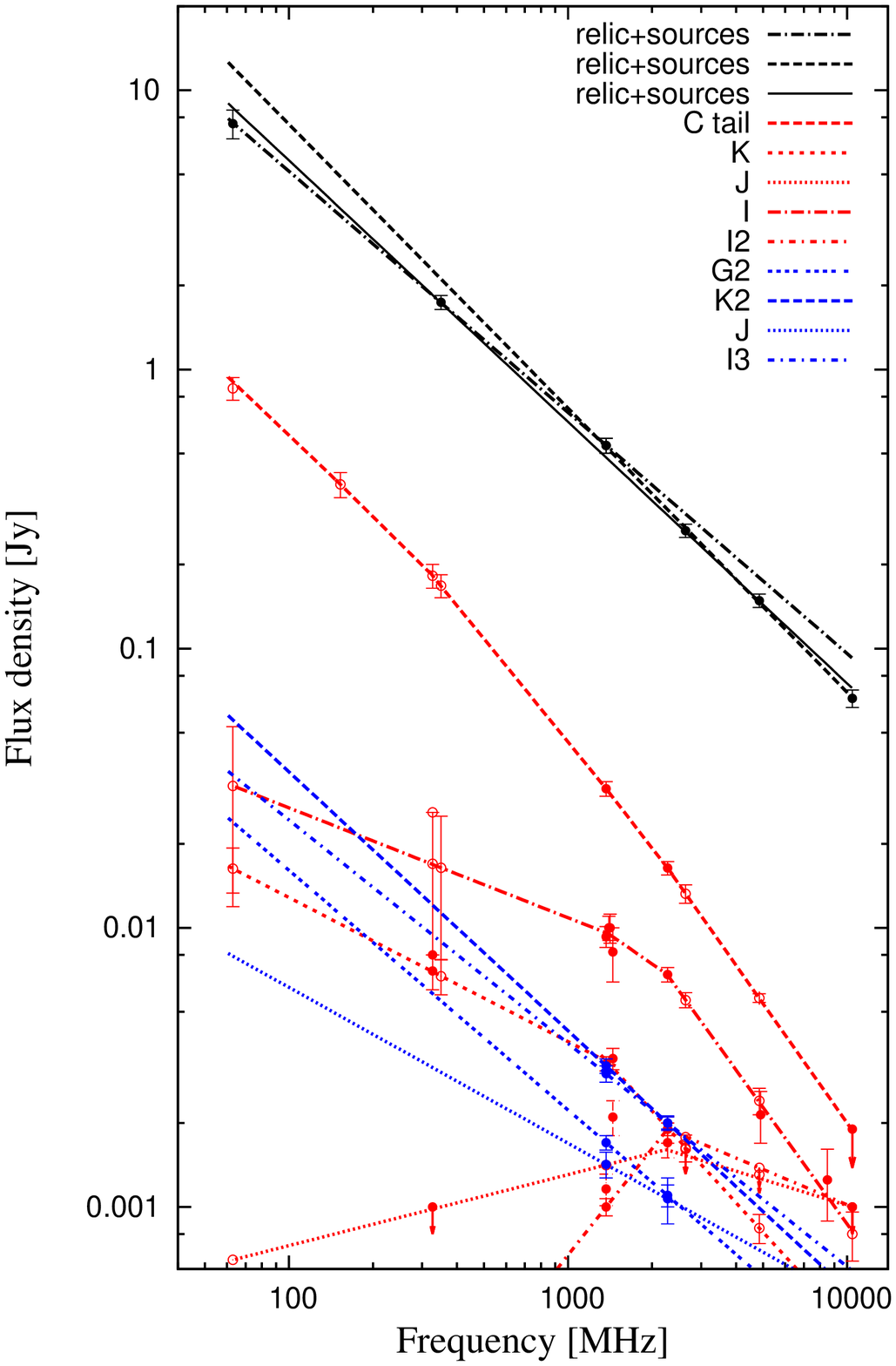}
     \caption{ Integrated radio spectra of the different components in the relic region shown in Fig. \ref{fig:spec_areas}. 
     In black we show the flux densities and spectra of the emission from the entire region (relic+sources). The dashed ($\alpha_{1369}^{10450}=1.02 \pm 0.02$) and dot-dashed 
     ($\alpha_{63}^{1369}= 0.86 \pm 0.01$) lines represent a double power-law fit to the data. 
     The solid line ($\alpha_{63}^{10450}= 0.93 \pm 0.02$) is a single power-law fit.
     In red and blue we show the spectra of the discrete sources included in the region. Red spectra are plotted individually in 
     Figs. from  \ref{fig:Cspe} to \ref{fig:I2spe}. 
     Blue spectra are straight power-law fits to the measured flux densities at 1369 and 2273 MHz. See text for more details. }
         \label{fig:relic+soursp}
   \end{figure}
Figure \ref{fig:spec_areas} shows the area selected for measuring the radio relic flux densities. 
From the high-resolution image it is possible to see which are the discrete sources included in such area: the tail of source C C$_{tail}$, K, J, I, G2, K2, J2, I2 and I3.
To estimate the flux density from the discrete sources that needs to be subtracted from the total emission, 
we produced two images at 1369 MHz (VLA C configuration) and 2273 MHz (WSRT) using the same uv-range (262-15460 $\lambda$), pixel-size and restoring beam and we calculated 
the spectral index of the discrete sources embedded in the relic emission between these two frequencies. 
Moreover we measured the flux density at 10450 MHz for the different components of source C and for the sources J and I2. 
Where available, we combined our data with data
 collected from the literature and we modeled the integrated spectra of the discrete sources over the frequency range 63 - 10450 MHz. 
 All the measured and extrapolated flux densities, as well as the spectral indices of the source C components with references are listed in Table 3. 
 Details on the flux densities derivation are given in Appendix \ref{app:sources}.
The spectra of the sources C, K, J, I and I2, for which more than two flux density measurements were available, are plotted individually in Figs. from \ref{fig:Cspe} to \ref{fig:I2spe} and all
together in red in Fig. \ref{fig:relic+soursp}.
 For the sources K2, J2, I3 and G2 only our flux density measurements at 1369 and 2273 MHz were available and we simply assumed straight spectra. This assumption may lead to a slight 
over estimate of the flux densities at both low and high frequencies (as for standard radio source synchrotron spectra we expect a 
flattening at low frequencies and a steepening at
high frequencies). On the other hand, they are weak sources and their flux densities are not crucial for the relic spectrum determination. 
Their spectra are plotted in blue in Fig. \ref{fig:relic+soursp}. This figure
shows the spectra of the discrete sources included in the region selected,
compared to the total flux density in the region (relic+sources, black lines).
The main flux density contribution among the discrete sources in the relic area come from the tail of source C, especially at lower frequencies. The source appears 
 noticeably narrow and long at high resolution, $\sim$ 410$\arcsec$ in total in our 2273 MHz image (see Fig. \ref{fig:spec_areas}).
 The published flux densities in the literature refer to the source as a whole. 
 We modeled the source distinguishing between the head (long 76$\arcsec$) and the tail as we are interested 
 only on the contribution from the latter. 
In Figure \ref{fig:Cspe} the spectrum over the range 63-10450 MHz is plotted for the entire source, the head, and the tail. 
 As common among head-tail sources, the head is flatter than the tail and the spectrum steepens at high frequency for both components.
 At lower frequencies, the tail contains almost all the emission of the sources. At high frequencies the electrons in the tail are older while the head is more clearly 
 visible.\\
 Sources J and I2 have an inverted-spectrum. Their spectra (plotted in Figs. \ref{fig:Jspe} and \ref{fig:I2spe}) have a convex shape at GHz frequencies, 
 typical of inverted-spectrum sources, and are likely young radio objects.\\
  \begin{sidewaystable*}[ht]
  \caption{Properties of the radio sources embedded in the radio relic emission.}\label{table:pointsources}
\centering 
\subfloat[Source C flux densities. \label{table:Cflux}]{
\begin{tabular}{c c c c c c c c c c } 
\hline
\hline
\noalign{\smallskip}
Component & 63 MHz  & 153 MHz   & 327 MHz   & 351 MHz & 1369 MHz & 2273 MHz & 2640 MHz & 4850 MHz & 10450 MHz\\
          & S (mJy)  & S (mJy)    & S (mJy)    & S (mJy)  & S (mJy)   & S (mJy)   & S (mJy)   & S (mJy)   & S (mJy)  \\
\noalign{\smallskip}
\hline
\noalign{\smallskip}
ALL  & {\itshape 1048 $\pm$ 256} \tablefootmark{a}       & 480 $\pm$ 50 \tablefootmark{b}                  & 246 $\pm$ 20 \tablefootmark{c}                & {\itshape 228.7 $\pm$ 17.7} \tablefootmark{a}  &56.5 $\pm$ 3.4 \tablefootmark{d}  & 33.7 $\pm$ 1.8 \tablefootmark{d}  & {\itshape 27.3 $\pm$ 1.4} \tablefootmark{a}   &{\itshape 11.7 $\pm$ 1.3} \tablefootmark{a}  & < 4.0 $\pm$ 0.9  \tablefootmark{d}\\
HEAD & {\itshape 144.5 $\pm$ 26.9} \tablefootmark{a}     & {\itshape 92.7$\pm$ 10.7} \tablefootmark{a}     &{\itshape 63.4 $\pm$ 5.5} \tablefootmark{a}    & {\itshape 60.6 $\pm$ 5.1 } \tablefootmark{a}   & 25.0 $\pm$ 1.5 \tablefootmark{d} & 17.3 $\pm$ 0.9 \tablefootmark{d}  & {\itshape 14.0 $\pm$ 0.8 } \tablefootmark{a}  & {\itshape 6.0 $\pm$ 0.9} \tablefootmark{a}  & 2.1 $\pm$ 0.6 \tablefootmark{d}\\
TAIL & {\itshape 903.5 $\pm$ 257.0 } \tablefootmark{a}   & {\itshape 387.3 $\pm$ 51.1 } \tablefootmark{a}  & {\itshape 182.6 $\pm$20.7 } \tablefootmark{a} & {\itshape 168.1 $\pm$ 18.4} \tablefootmark{a}  &31.5 $\pm$ 1.9 \tablefootmark{d}  & 16.4 $\pm$ 0.9 \tablefootmark{d}  & {\itshape 13.3 $\pm$ 1.6} \tablefootmark{a}   & {\itshape 5.7 $\pm$1.5 } \tablefootmark{a}  & < 1.9 $\pm$ 0.7\tablefootmark{d}\\
\noalign{\smallskip}
\hline
\end{tabular}
}
\vspace{0.5cm}
\subfloat[Source C spectral indices. \label{table:Cindex}]{
\begin{tabular}{l c c c c c } 
\hline
\hline
\noalign{\smallskip}
Component & $\alpha _{63}^{153}$ & $\alpha _{153}^{327}$ & $\alpha _{327}^{1369}$ & $\alpha _{1369}^{2273}$ & $\alpha _{2273}^{10450}$\\
\noalign{\smallskip}
\hline
\noalign{\smallskip}
ALL  & {\itshape 0.88 $\pm$ 0.17} \tablefootmark{a} & 0.88 $\pm$ 0.17 \tablefootmark{d}            & 1.03 $\pm$ 0.07 \tablefootmark{d}             & 1.02 $\pm$ 0.16 \tablefootmark{d} & 1.40 $\pm$ 0.14 \tablefootmark{d} \\ 
HEAD & {\itshape 0.50 $\pm$ 0.10} \tablefootmark{a} & {\itshape 0.50 $\pm$ 0.10} \tablefootmark{e} & {\itshape 0.65 $\pm$ 0.10 } \tablefootmark{f} & 0.72 $\pm$ 0.15 \tablefootmark{d} & 1.40 $\pm$ 0.19 \tablefootmark{d} \\
TAIL & {\itshape 0.95 $\pm$ 0.35} \tablefootmark{a} & {\itshape 1.00 $\pm$ 0.22} \tablefootmark{a}  &{\itshape 1.23 $\pm$ 0.10 } \tablefootmark{a} & 1.30 $\pm$ 0.16 \tablefootmark{d} & {\itshape 1.42 $\pm$ 0.25} \tablefootmark{d} \\
\noalign{\smallskip}
\hline
\end{tabular}
}
\vspace{1cm}

\centering
\subfloat[Other sources flux densities. \label{table:sources_flux}]{
\begin{tabular}{ c c c c c c c c c c c c} 
\hline
\hline
\noalign{\smallskip}
Source &63 MHz                     &327 MHz  & 351 MHz                   & 1369 MHz                   &1446 MHz                   &2273 MHz                   &2640 MHz                   & 4850 MHz                  & 4900 MHz                & 8500 MHz                & 10450 MHz\\
       &S(mJy) \tablefootmark{a}   &S(mJy)   &  S(mJy) \tablefootmark{a} &  S(mJy) \tablefootmark{d}  &  S(mJy) \tablefootmark{c} &  S(mJy) \tablefootmark{d} &  S(mJy) \tablefootmark{a} &  S(mJy) \tablefootmark{a} & S(mJy) \tablefootmark{g}& S(mJy) \tablefootmark{g}& S(mJy)\\
\noalign{\smallskip}
\hline
\noalign{\smallskip}
K  & {\itshape 16.3 $\pm$ 3.0}    & 7.0 $\pm$ 1.0 \tablefootmark{c}                & {\itshape 6.7 $\pm$ 0.1 }   &  3.3 $\pm$ 0.2    &3.4 $\pm$ 0.3  & 1.9 $\pm$ 0.2 & {\itshape 1.6 $\pm$ 0.2 }  &{\itshape 0.8 $\pm$ 0.1 }  &(...)         & (...)         &{\itshape $\ll$1} \tablefootmark{a} \\
J  & {\itshape < 0.65 }           & < 1 \tablefootmark{c}                          &{\itshape  1.0 $\pm$ 0.1 }   &  1.2 $\pm$ 0.1    & 2.1 $\pm$ 0.3 & 1.7 $\pm$ 0.2 & {\itshape < 1.5  }         & {\itshape < 1.2  }        & (...)        & (...)         & < 1 \tablefootmark{d} \\
I  & {\itshape  32.2 $\pm$ 20.3 } & {\itshape17.0 $\pm$ 8.9 } \tablefootmark{a}    & {\itshape16.4 $\pm$ 8.7 }   & 9.3 $\pm$ 0.8     & 8.2 $\pm$ 1.8 & 6.8 $\pm$ 0.4 & {\itshape 5.5 $\pm$ 0.3}   & {\itshape2.4 $\pm$ 0.3 }  &2.1 $\pm$ 0.5 & 1.2 $\pm$ 0.4 &{\itshape0.8 $\pm$ 0.2} \tablefootmark{a}\\
I2 & {\itshape  $\ll$1 }          &{\itshape  $\ll$1 } \tablefootmark{a}           & {\itshape  $\ll$1 }         & 1.0 $\pm$ 0.07    & (...)         & 1.9 $\pm$ 0.1 & {\itshape < 1.8 }          & {\itshape < 1.4 }         &(...)         &(...)          & < 1 \tablefootmark{d}\\
K2 & {\itshape 56.0 $\pm$ 27.8 }  & (...)                                          & {\itshape  11.3 $\pm$ 2.6 } & 3.2 $\pm$ 0.19    & (...)         & 2.0 $\pm$ 0.1 & {\itshape 1.7 $\pm$ 0.1 }  &{\itshape 1.0 $\pm$ 0.1 }  &(...)         & (...)         &{\itshape  $\ll$ 1} \tablefootmark{a} \\
J2 & {\itshape 7.9 $\pm$6 }       & (...)                                          & {\itshape 3.0 $\pm$ 1.0 }   & 1.4 $\pm$ 0.2     & (...)         & 1.1 $\pm$ 0.2 & {\itshape 1.0 $\pm$ 0.1 }  & {\itshape 0.7 $\pm$ 0.1 } &(...)         &(...)          &{\itshape $\ll$1} \tablefootmark{a}\\
I3 & {\itshape  35.2 $\pm$ 18.6 } &  (...)                                         &{\itshape 8.9 $\pm$ 2.1 }    &3.0 $\pm$ 0.2      & (...)         & 2.0 $\pm$ 0.1 & {\itshape 1.8 $\pm$ 0.1 }  &{\itshape 1.1 $\pm$ 0.1 }  &(...)         & (...)         &{\itshape $\ll$ 1} \tablefootmark{a} \\
G2 & {\itshape  23.9 $\pm$ 13.7 } & (...)                                          &  {\itshape 5.5 $\pm$ 1.4  } & 1.7 $\pm$ 0.1     & (...)         & 1.1 $\pm$ 0.1 & {\itshape  1.0 $\pm$ 0.1 } & {\itshape  0.6 $\pm$0.1 } &(...)         &(...)          & {\itshape  $\ll$1} \tablefootmark{a}\\
\noalign{\smallskip}
\hline
\end{tabular}
}
\tablefoot{ Values in {\itshape italic} are assumed or extrapolated in this work, while values in normal font are measured values either in this work or published in 
the literature. For more details consult references below and text. \\
\tablefoottext{a}{Extrapolated in this work.}\\
\tablefoottext{b}{Measured by\citet{2012A&A...543A..43V}.}\\
\tablefoottext{c}{Measured by \citet{1994ApJ...436..654R}.}\\
\tablefoottext{d}{Measured in this work.}\\
\tablefoottext{e}{Deduced from Fig. 5.4 {\itshape Right site}, \citet{2009PhDT.........1I}.}\\
\tablefoottext{f}{Deduced from Fig. 7 {\itshape Right site}, \citet{1994ApJ...436..654R}}\\
\tablefoottext{g}{Measured by \citet{2009ApJ...694..992L}.}\\
          }   
\end{sidewaystable*}

\subsubsection{Radio relic spectrum}\label{subsec:relic_spec}
To avoid that resolution effects at different frequencies may alter the determination of the relic integrated spectrum, we convolved 
all the images used for the flux densities computation to the same resolution of $4\farcm4 \times$ 4$\farcm4\ $. At such low resolution, it is not easy to identify 
the region where to measure the relic flux density. We adopted the following strategy:
we first computed the relic flux density from the 4850 MHz image
at full resolution, where it is easier to separate the relic emission from the complex A+B emission. 
We then choose the relic region in the 4850 MHz image convolved to the resolution of 4$\farcm$4 $\times$ 4$\farcm$4, matching the flux density measured at full resolution at the same frequency.
The selected area is shown in Fig. \ref{fig:spec_areas}.
The region is further divided into two regions of enhanced radio brightness (regions G and H) discussed in the next section. 
As discussed in the previous section, this region include the radio relic and sources 
C$_{tail}$, K, J, I, G2, K2, J2, I2 and I3.\\
We first considered the total emission from the region.
Although the flux density measurements of the relic+sources in the range 63-10450 MHz can be fitted with a single power law with $\alpha_{63}^{10450}= 0.93 \pm 0.02$ 
(Fig. \ref{fig:relic+soursp}), hints of a steepening at frequencies > 1400 MHz are present. A separate fit of the
spectra between 63 and 1369 MHz and between 1369 and 10450 MHz shows indeed that these two frequency ranges are best represented by two different power laws, with 
$\alpha_{63}^{1369}= 0.86 \pm 0.01$ and $\alpha_{1369}^{10450}=1.02 \pm 0.02$ (Fig. \ref{fig:relic+soursp}).
All the fits are plotted over the entire range 63-10450 MHz to highlight differences.
Since the relic is the major contributor in the region both at low and high frequency,
this suggests that a steepening might be present in its spectrum as well.
Moreover the relic spectrum between 63 and 1369 MHz must be $\alpha(relic)_{63}^{1369}\leq 0.86\pm 0.01$ as it cannot be steeper than the 
relic+sources spectrum. Similarly it must be  $\alpha(relic)_{1369}^{10450}\geq 1.02\pm 0.02$ as it cannot be flatter than the relic+sources spectrum.
Indeed, after discrete sources subtraction we find that, although the relic spectrum between 63 and 10450 MHz is not inconsistent with a single power law with 
$\alpha(relic)_{63}^{10450}= 0.92\pm 0.02$, it is 
best represented by two different power laws, with 
$\alpha(relic)_{63}^{1369}= 0.85\pm 0.01$ and $\alpha(relic)_{1369}^{10450}= 1.00\pm 0.02$ 
(Fig. \ref{fig:relic_sp}). This is supported by a lower values of the reduced $\chi^2$ in the case of the double power law ($\chi^2_{red}=0.19$)
respect to the single power law ($\chi^2_{red}=0.92$).
The measured relic flux densities (before and after discrete sources subtraction) are summarized in Table \ref{table:relicflux}.
The flux densities measured at 63 MHz, 351 MHz and 1369 MHz are in agreement within the error bars with the already published
flux densities at the same frequencies. The low-frequency spectral index is in agreement with what was found by \citet{2012A&A...543A..43V}.\\
\citet{2006AJ....131.2900C} report a relic mean spectral index between 1369 MHz and 1703 MHz of 1.2. 
However, this value has a big uncertainty (not quoted by author) since it is derived as the mean value from the spectral index image between two very close frequencies.
We cannot therefore exclude it is in agreement with our value. 
Moreover, the very recent JVLA observations by \citet{2014arXiv1408.5931O} reports for the relic an overall intensity weighted spectral index in the L-band
of $\sim$ 0.94, in agreement with our finding.
\subsubsection{Regions G and H}   
\begin{figure*}[!ht]
\centering
\subfloat[Total cluster.]{
\label{fig:totclustsp}
\includegraphics[trim=0cm 5.5cm 2cm 4cm, clip=true, width=9.5cm]{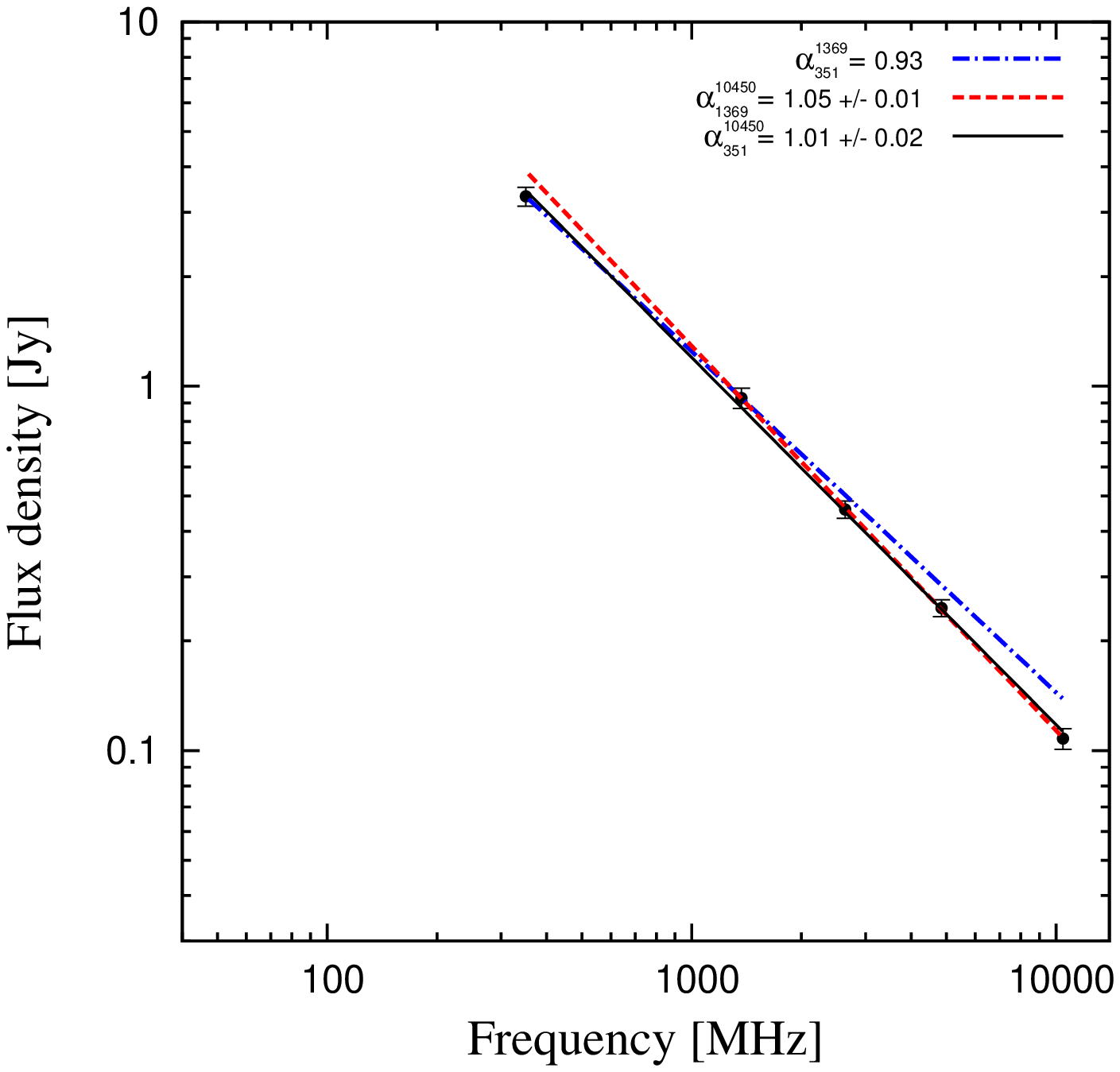}}
\subfloat[Radio relic.]{
\label{fig:relic_sp}
\includegraphics[trim=0cm 5.5cm 2cm 4.5cm, clip=true, width=9.5cm]{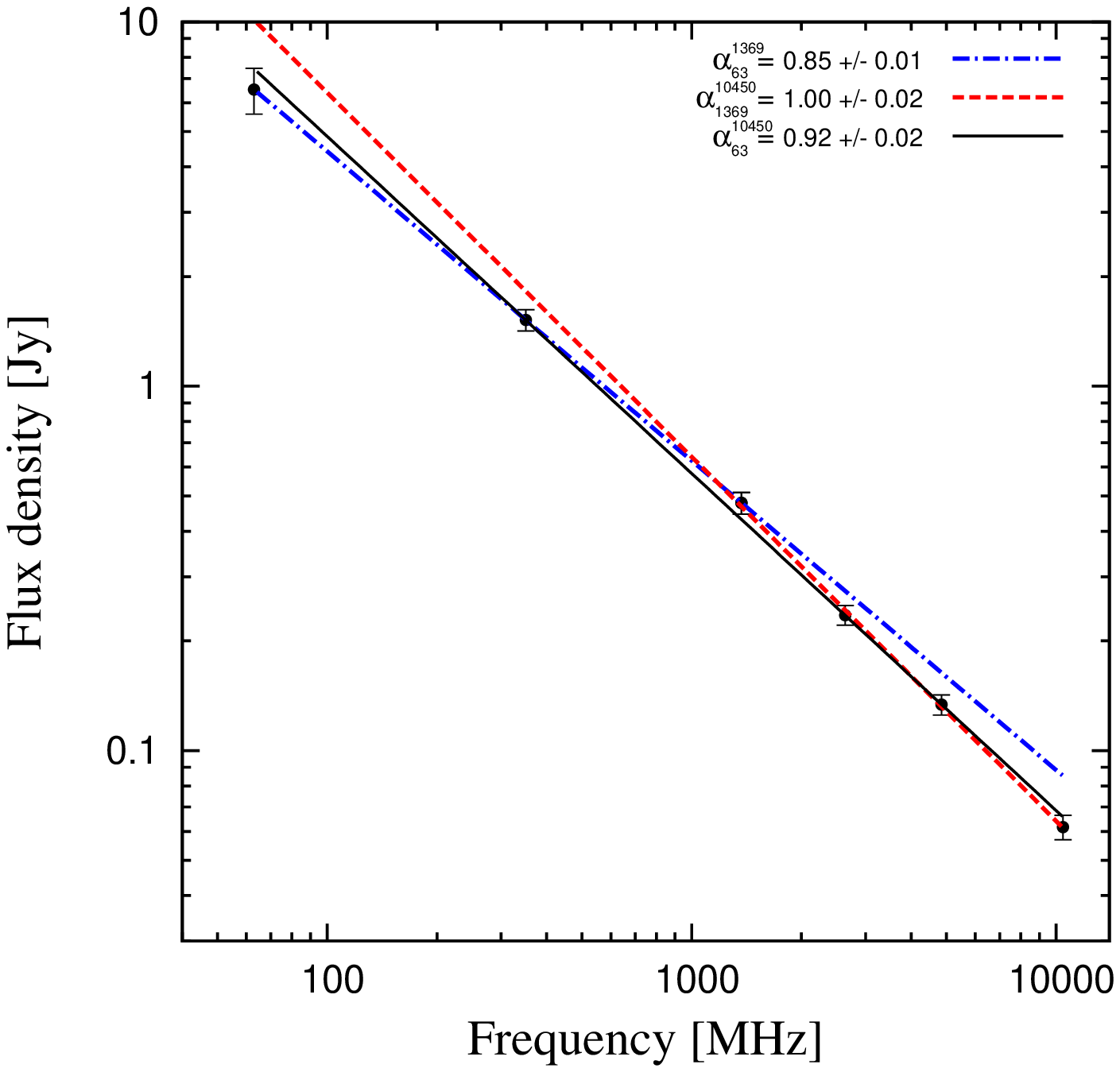}}\\
\subfloat[G region.]{
\label{fig:Gsp}
\includegraphics[trim=0cm 6.5cm 2cm 6.5cm, clip=true, width=9.5cm]{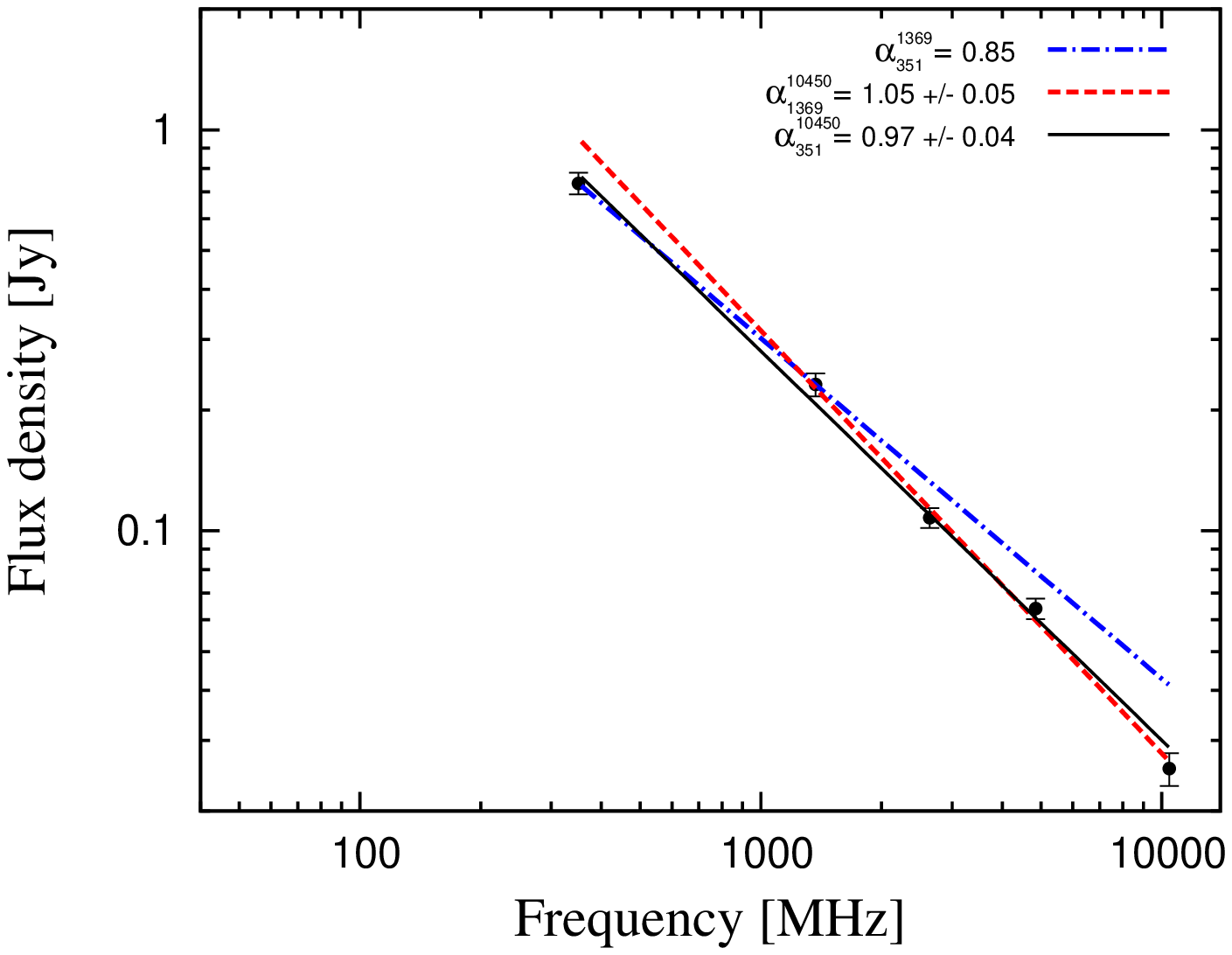}}
\subfloat[H region.]{
\label{fig:Hsp}
\includegraphics[trim=0cm 6.5cm 2cm 7cm, clip=true, width=9.5cm]{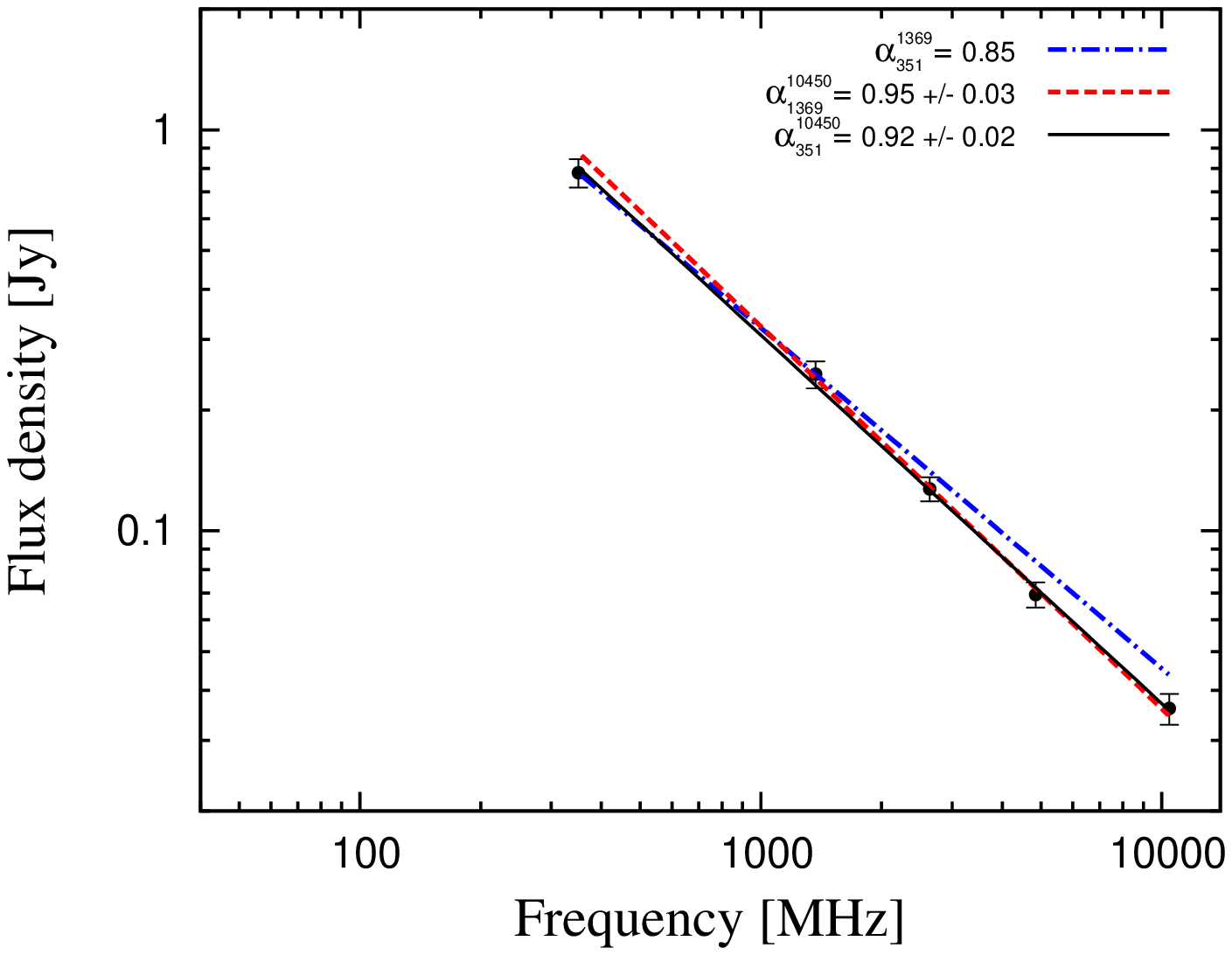}}
\caption{Integrated radio spectra. The red dashed and blue dot-dashed lines are a double power-law fit to the data while the solid black line is the single power-law fit.
In panel (b) the green crosses are the upper limits to the radio relic flux taking into account the SZ effect.
See text for more details.}
\label{fig:spectra}
\end{figure*}

Our high-resolution image shows that the relic can be divided into two separate parts: regions G and H in Fig. \ref{fig:spec_areas}.
The cases of double relics in the same cluster are getting more and more common since the first discovery of two almost
symmetric relics located on opposite sides in A3667. Since then, several other double relics systems have been found (see Feretti et al. 2012). 
We have investigated the possibility that the two different parts have different properties fitting the spectra in the frequency range 351-10450 MHz, separately.
After discrete sources subtraction, the flux densities of region G can be fitted with a single power law with a spectral index of
$\alpha(G)_{351}^{10450}=0.97\pm0.04$ (Fig. \ref{fig:Gsp}).
For homogeneity with the radio relic analysis, we performed a separate fit of the spectra between 351 and 1369 MHz and between 1369 and 10450 MHz.
We find $\alpha(G)_{351}^{1369}= 0.85$ and $\alpha(G)_{1369}^{10450}= 1.05\pm 0.05$
(Fig. \ref{fig:Gsp}).\\
For the H region the flux densities can be modeled with
a single power law with spectral index $\alpha(H)_{351}^{10450}=0.92 \pm 0.02$  
or with a double law with the same slope as the G region at low frequency 
$\alpha(H)_{351}^{1369}= 0.85$ and a bit flatter spectra respect the G region at high frequency $\alpha(H)_{1369}^{10450}= 0.95\pm 0.03$
(Fig. \ref{fig:Hsp}).\\
The measured flux densities before and after discrete sources subtraction are listed in Tables \ref{table:Gflux} and \ref{table:Hflux}, 
while the spectra are plotted in Figs. \ref{fig:Gsp} and \ref{fig:Hsp}.

\begin{table*}[!ht]
\begin{minipage}[t]{.5\textwidth}
\vspace{0pt}
\centering 
\captionof{table}{Total cluster flux densities.}
\label{table:totflux}
\begin{tabular}{c c} 
\hline
\hline
\noalign{\smallskip}
$\nu$ (MHz) & S(mJy)          \\
\noalign{\smallskip}
\hline
\noalign{\smallskip}
351         & 3320.0 $\pm$ 200.0  \\
1369        & 928.6  $\pm$ 57.0 \\
2640        & 459.0 $\pm$ 24.8  \\ 
4850        & 246.3 $\pm$ 13.3  \\
10450       & 107.8 $\pm$ 7.5  \\
\noalign{\smallskip}
\hline
\end{tabular}
\vspace{0pt}
\end{minipage}
\begin{minipage}[t]{.5\textwidth}
\vspace{0pt}
\centering
\vspace{0pt}
\captionof{table}{Radio relic flux densities.}
\label{table:relicflux}
\begin{tabular}{c c c} 
\hline
\hline
\noalign{\smallskip}
$\nu$ (MHz) & S(mJy) & S(mJy)         \\
            & Before subtraction& After subtraction\\
\noalign{\smallskip}
\hline
\noalign{\smallskip}
63     &7600.0 $\pm$ 900.0     & 6520.0 $\pm$ 940.0   \\
351    &1740.0 $\pm$ 100.0     & 1520.0 $\pm$ 100.0 \\
1369   &534.3 $\pm$ 33.1    & 478.5 $\pm$ 33.2   \\ 
2640   &264.6 $\pm$ 14.4     & 235.4 $\pm$ 14.5   \\
4850   &148.6 $\pm$ 8.2   & 133.7 $\pm$ 8.5  \\
10450  &66.4 $\pm$ 4.8  & 61.7 $\pm$ 4.8 \\ 
\noalign{\smallskip}
\hline
\end{tabular}
\end{minipage}
\end{table*}

\begin{table*}[!ht]
\begin{minipage}[t]{.5\textwidth}
\vspace{0pt}
\centering
\vspace{0pt}
\captionof{table}{Region G flux densities.}
\label{table:Gflux}
\begin{tabular}{c c c}
\hline
\hline
\noalign{\smallskip}
$\nu$ (MHz) & S(mJy) & S(mJy)         \\
            & Before subtraction& After subtraction\\
\noalign{\smallskip}
\hline
\noalign{\smallskip}
351    &741.2 $\pm$ 45.8 &735.7 $\pm$ 45.8  \\
1369   &233.3 $\pm$ 15.1 &231.6 $\pm$ 15.1   \\ 
2640   &108.8 $\pm$ 6.2 &107.8 $\pm$ 6.2   \\
4850   &64.6 $\pm$ 3.8  &64.0  $\pm$ 3.8   \\
10450  &25.5 $\pm$ 2.4 &25.5  $\pm$ 2.4 \\ 
\noalign{\smallskip}
\hline
\end{tabular}
\vspace{0pt}
\end{minipage}
\begin{minipage}[t]{.5\textwidth}
\vspace{0pt}
\centering
\vspace{0pt}
\captionof{table}{Region H flux densities.}
\label{table:Hflux}
\begin{tabular}{c c c}
\hline
\hline
\noalign{\smallskip}
$\nu$ (MHz) & S(mJy) & S(mJy)         \\
            & Before subtraction& After subtraction\\
\noalign{\smallskip}
\hline
\noalign{\smallskip}
351    &996.8 $\pm$ 60.9 &781.3 $\pm$ 64.3  \\
1369   &299.6 $\pm$ 19.0 &245.8  $\pm$ 19.1   \\ 
2640   &155.5 $\pm$ 8.6  &127.2 $\pm$ 8.8 \\
4850   &83.7 $\pm$ 4.8   &69.4 $\pm$ 5.0   \\
10450  &40.7 $\pm$ 3.1  &36.0 $\pm$ 3.2 \\ 
\noalign{\smallskip}
\hline
\end{tabular}
\end{minipage}
\end{table*}

\begin{table*}[!ht]
\caption{Observed synchrotron spectral indices of the different components.} 
\centering 
\begin{tabular}{c c c c} 
\hline
\hline
\noalign{\smallskip}
 & Single power-law fit &\multicolumn{2}{c}{Double power-law fit}\\
 \hline
 \vspace{4pt}
                   &$\alpha_{63}^{10450}$ & $\alpha_{63}^{1369}$ & $\alpha_{1369}^{10450}$\\
Total relic region & 0.93 $\pm$ 0.02      & 0.86 $\pm$ 0.01    &1.02 $\pm$ 0.02         \\
Radio relic        & 0.92 $\pm$ 0.02      & 0.85 $\pm$ 0.01  & 1.00 $\pm$ 0.02        \\
\hline
\vspace{4pt}
                   &$\alpha_{351}^{10450}$& $\alpha_{351}^{1369}$ & $\alpha_{1369}^{10450}$\\
Total cluster      &1.01 $\pm$ 0.02       & 0.93                  & 1.05 $\pm$ 0.01\\
Region G           &0.97 $\pm$ 0.04       &0.85                   & 1.05 $\pm$ 0.05 \\
Region H           &0.92 $\pm$ 0.02       &0.85                   & 0.95 $\pm$ 0.03\\
\hline
\end{tabular}
\label{table:alphas} 
\end{table*}

\section{X-ray analysis and results}\label{sec:xray_results}
\subsection{ICM temperature}\label{subsec:Treg}
\begin{figure}[!ht]
   \centering
   \includegraphics[width=8.5cm]{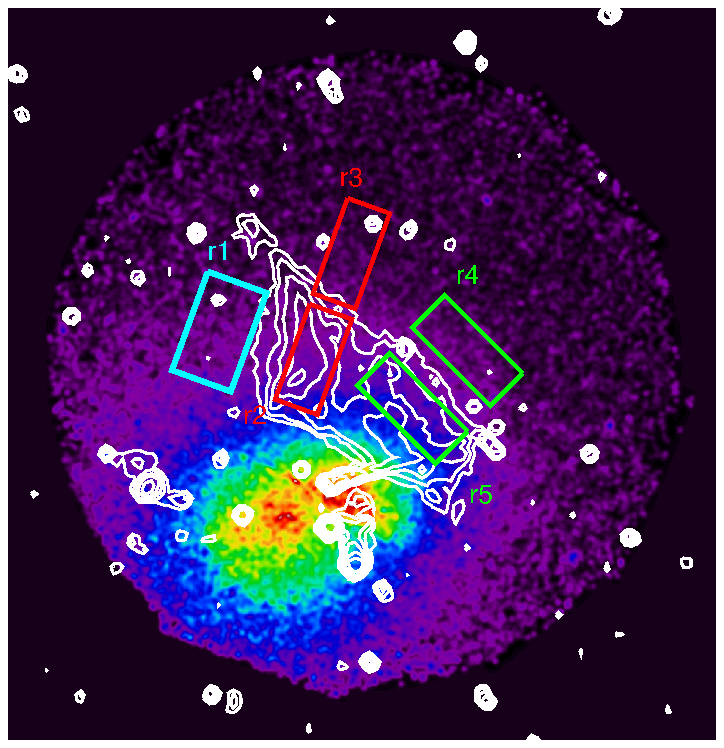}
     \caption{Regions used for the ICM temperature extraction. Colors show the XMM-Newton X-ray image, while the contours show the radio emission at a lower resolution respect to 
     Fig. \ref{fig:WSRT_TI}.}
         \label{fig:xmm}
   \end{figure}

In the course of a merger, a significant portion of the energy involved is dissipated by shocks and turbulence leading eventually to the heating of the ICM gas.
As mentioned in the Introduction, in the test particle regime of DSA theory 
the shock structure is determined by the canonical shock jump conditions (Rankine-Hugoniot).
Applying these conditions, assuming the ratio of specific heats as 5/3,
the expected ratio between the postshock and preshock temperatures, respectively $T_2$ and $T_1$, is related to the shock Mach number through the following relation:
\begin{equation}\label{eq:Tratio}
\frac{T_2}{T_1}=\frac{5M^4+14M^2-3}{16M^2} .
\end{equation}
We extracted the ICM temperature in different 
regions across the radio relic corresponding to the rectangular areas shown in Fig. \ref{fig:xmm}.
The observed spectra were assumed to consist of thin thermal plasma
emission from the ICM, plus the total background contamination described in Sec. \ref{sec:xray_obs}.
The emission from the ICM was modeled with an additional absorbed thermal component in the total model (ICM+background).
For the Suzaku analysis, to generate the auxiliary response files (ARF), we used an image constructed using a 
$\beta$-model \citep[$\beta=0.816$, $r_c=5.64^\prime$ from][]{1997ApJ...491..467M} as input for the $xissimarfgen$.
For XMM-Newton data, we fitted the spectra in the 0.5-8 keV energy range, excluding the 1.4-1.6 keV 
band due to the strong contamination from the Al line in all three detectors.
Because of the low number of counts in the XMM-Newton spectra of region r1, r2 and r3, we kept the metallicity values in the fit frozen to the value obtained with the Suzaku 
analysis, which are better constrained \footnote{(Z(r1)= $0.27 \pm 0.05$ Z$_{\odot}$, Z(r2)= $0.12 \pm 0.04$ Z$_{\odot}$, Z(r3)= $0.19 \pm 0.12$ Z$_{\odot}$).}.
Temperatures and normalizations of the thermal components were allowed to vary in the fit.\\
We observe a temperature drop across the G region of the radio relic, with the temperature jumping from 8.45$_{-0.33}^{+0.43}$ keV in the region r2 (on the relic) to 
4.95$_{-0.39}^{+0.48}$ keV in the region r3 (outside the outer edge of the relic). 
XMM-Newton measurements in the same regions provide consistent temperatures although with bigger errors. 
This is in agreement with \citet{2002ApJ...565..867S} that found a hot region 
($\sim$ 9 KeV) in positional coincidence with the G lobe of the radio relic.
Radio relics are usually associated to outgoing merger shocks that travels from the core of a merging event outwardly towards the periphery of the clusters.
If we apply such scenario to A\,2256, connecting the radio relic emission to a shock front propagating 
outwardly in the north-western direction across the G region of the radio relic, we can consider r2 as the post-shock region and r3 as the preshock region.
In this case we have a temperature ratio $T_2/T_1$=1.7. It is important to notice that this estimate is impacted by the angle of the shock front to the line of sight.
If the shock is not in the plane of the sky, as it is likely the case for A\,2256, projected mixing of shocked hot gas with cool gas will reduce the temperature
of the shocked gas and increase the temperature of the cool gas, basically dropping the temperature ratio.\\
Regions r5 and r4, respectively on the H region of the relic and outside the outer edge, show almost equal temperatures $T(r5)$=5.89$_{-0.70}^{+0.95}$ and 
$T(r4)$=5.89$_{-1.13}^{+1.06}$ in the XMM-Newton images, although with big uncertainties. Unfortunately, region r4 is out of
Suzaku field-of-view and cannot be checked. This is anyway in agreement with previous studies 
conducted by \citet{2002ApJ...565..867S} and \citet{2008A&A...479..307B} that showed that this region is cold due to the presence of a cool
subcluster at an early stage merging with the main cluster, approaching from somewhere west. 
Projection effects, affecting in particular region r5, can thus be responsible for the non detection of a temperature jump across the H region of the relic.\\
The measured temperatures in the different regions are summarized in Table \ref{table:Tregions}
for both Suzaku and XMM-Newton data.\\
\begin{table*}
\caption{ ICM temperatures kT(keV).} 
 \centering 
\begin{tabular}{c c c} 
\hline
\hline
\noalign{\smallskip}
Region & Suzaku data & XMM-Newton data \\
\noalign{\smallskip}
\hline
\noalign{\smallskip}
\vspace{4pt}
r1    & 6.53$_{-0.23}^{+0.28}$ & 6.70$_{-1.07}^{+1.04}$  \\
\vspace{4pt}
r2 (on G region)   & 8.45$_{0.33}^{+0.43}$  & 8.29$_{-0.88}^{+1.16}$\\
\vspace{4pt}
r3    & 4.95$_{-0.39}^{+0.48}$ & 4.97$_{-1.83}^{+2.11}$\\
\vspace{4pt}
r4    & (...)                    & 5.89$_{-1.13}^{+1.06}$\\
r5 (on H region)   & (...)                    & 5.89$_{-0.70}^{+0.95}$\\
\noalign{\smallskip}
\hline
\end{tabular}
\label{table:Tregions} 
\end{table*}
\section{Is the Sunyaev-Zeldovich effect important?}\label{Sec:SZ}
The thermal Sunyaev-Zeldovich (SZ) effect \citep{1970CoASP...2...66S} consists in the inverse Compton scattering to higher energies of the photons of the Cosmic Microwave Background (CMB) 
that interact with the hot electrons in the ICM of galaxy clusters. 
The effect depends on the pressure produced by the plasma along the line of sight and is parametrized through the Comptonisation parameter
$y$ \citep[see][for a review]{2002ARA&A..40..643C}.
The SZ effect produces a modification of the CMB blackbody spectrum,
creating a negative flux bowl on the scale of the cluster, at GHz frequencies.
This might lead to two sources of errors in our flux densities measurements at high frequencies.
The first derives from a wrong setting of the zero-level in the single dish images.
The zero-level in single dish observations is usually set assuming 
the absence of sources at the map edges, setting the intensity level to zero there, and interpolating linearly between the two opposite edges of the map.
If the SZ decrement is important in the regions where zero flux was assumed 
(determining a negative flux in those areas), this results in a wrong setting of the zero-level of the entire map. This lead to a lower estimation of the flux densities 
integrated over any region in the map. 
In addition to this, there is the SZ decrement specific to the area used to compute the relic flux densities.\\
We used the integrated Comptonisation parameter $Y_{SZ}$ as measured by Planck \citep{2011A&A...536A..11P} 
and the universal pressure profile shape as derived by \citet{2010A&A...517A..92A}, to
derive the predicted pressure and Comptonisation parameter profiles for A\,2256 \citep[see][]{2013A&A...550A.131P}.
We used the derived $y$ parameter radial profile to 
estimate the importance of the SZ decrement in different regions at the edges of the original Effelsberg 10450 MHz image, to check the zero-level setting. 
Integrating on areas free of sources at the map edges with angular diameter of 4$^\prime$, we find values of -0.1 / -0.3 mJy for the CMB flux decrement. We conclude that 
the effect of the SZ decrement on the zero-level setting of the 10450 MHz image is negligible. As the SZ effect increase increasing the frequency, we conclude that 
it does not affect the zero-level setting of the 2640 and 4850 MHz Effelsberg images.\\
Estimating the second effect is more complicated, especially if we believe that the relic in A\,2256 is powered by a shock.  
In case of a shock, in fact, a sharp increase of the pressure at the shock front is expected, causing a local increase of the SZ decrement.
An exact estimate of the amount of the effect requires a detailed knowledge of the shock geometry and orientation and of the pressure changes across 
the relic width. We estimate here an upper limit for the SZ effect on our high-frequencies flux estimates.
The radio relic fluxes have been calculated in an approximatively rectangular area of $\sim$ 18$^\prime \times$ 9$^\prime$, whose major axis is 
located at a projected distance of $\sim$ 7.2$^\prime$ ($\sim$ 500 kpc) from
the cluster X-ray main peak. 
If we assume that
we observe the shock in the plane of the sky, we can consider a physical distance of the relic from the cluster center of$\sim$ 500 kpc. This is a lower limit to the real 
distance relic-cluster center and maximize the SZ effect, that is higher in the denser central regions.
We can derive a qualitative estimation of the possible shock pressure jump from the observed ICM temperature ratio across the relic.
Inverting eq. $\ref{eq:Tratio}$ 
we obtain a Mach number $\sim$ 1.7. As discussed, this number can have been reduced by projection effects. We assume for the calculations a Mach number $M=2$. 
Applying the Rankine-Hugoniot jump conditions, the ratio between
the postshock $P_2$ and preshock $P_1$ pressures, given the Mach number $M$, is $P_2/P_1 \sim$4. 
We estimated the SZ flux decrement in the area used for the radio relic flux measurements using eq. 3 from \citet{2002ARA&A..40..643C} and assuming a value $y_{POST}=4y_{PRE}$ 
(where $y_{PRE}$ was extracted from the y-profile at a corresponding distance of 7.2$^\prime$) constant over all the area.
This assumptions bring to an over estimation of the SZ effect because the real width of the relic is expected to be much lower than the projected one, and the increased 
pressure is expected to decrease again moving away from the shock front in the downstream region.
With these assumptions, we obtain SZ flux decrements of $\sim$ -20 mJy, $\sim$ -4.3 mJy, $\sim$ -0.3 mJy respectively at 10450 MHz, 4850 MHz and 2640 MHz. 
We stress that this numbers are upper limits for the SZ decrement. Assuming, for example, the physical distance of $\sim $700 Kpc from the center deduced by  
\citet{1998A&A...332..395E} based on polarization properties, the effect at 10450 MHz reduces to $\sim$ -11 mJy.\\
We plotted in Fig. \ref{fig:relic_sp} (green crosses) the upper limits of the relic flux at 4850 and 10450 MHz, adding the upper limit SZ flux decrements (in absolute value) 
to the measured fluxes. 
The effect on the integrated radio spectrum is to make it even flatter. 
\section{Discussion}\label{sec:disc}
Independent of the acceleration mechanism, we can infer the range of energies (in terms of the Lorentz factor $\gamma_e$) of the emitting electrons 
in the relic region from the integrated spectrum via \citep{2012ApJ...756...97K}:
\begin{equation}
 \gamma_e \approx 1.26 \times 10^4 \left(\frac{\nu_{obs}}{1 GHz}\right)^{1/2} \left(\frac{B}{5 \mu G}\right)^{-1/2} (1+z)^{1/2}.
 \label{eq:gamma}
\end{equation}
The relic in A\,2256 has been observed from very low frequencies \citep[$\sim$ 20 MHz with LOFAR,][flux densities not published]{2012A&A...543A..43V} up to very high frequencies ($\sim$ 10 GHz with Effelsberg). 
This implies, via Eq. \ref{eq:gamma}, that the emitting electrons have energies between, at least,
$\gamma_{e,min} \sim 2\cdot 10^3 - 4 \cdot 10^3$ and $\gamma_{e,max} \sim 4\cdot 10^4 - 9 \cdot 10^4$. For both $\gamma_{e,min}$  and
$\gamma_{e,max}$, the two values refer to assumed magnetic fields $B = 5 - 1 \mu $G, respectively. 
The energy losses of low-energy electrons can be dominated by Coulomb interactions with the plasma, causing a low-frequency flattening of the integrated spectrum,
in the case that the density in the relic region is rather high and the strength of the magnetic field is low \citep{1999ApJ...520..529S}. 
The double power law found for the relic in A2256 can be looked at as due to a low-frequency flattening other than a high-frequency steepening.
However, assuming an electron density
of 10$^{-3}$ cm$^{-3}$ as estimated by \citet{1997ApJ...491..467M} at the relic position in A2256, and a typical value of 1 $\mu$G for the magnetic field, we calculated 
\citep[from eqs. 7 and 9 in][]{1999ApJ...520..529S}
that Coulomb losses do not dominate over synchrotron losses even for the lowest energy electrons in our range ($\gamma \sim 10^3$). 
The radiative lifetime for the less energetic electrons is $t_{rad}(\gamma_{e,min}) \sim 0.32 - 0.43$ Gyr 
(for $B = 5 - 1 \mu $G respectively) and $t_{rad}(\gamma_{e,max}) \sim 0.016 - 0.019$ Gyr for the more energetic ones.
Probably, whatever it is the acceleration mechanism, it is still at work since the particles emitting at 10 GHz lose rapidly their energy and the relic 
would become invisible at this frequency without a constant supply of new particles, soon after the end of the acceleration.
It is then reasonable to consider, in case of DSA, a continuous injection model with radiative losses dominated by IC and synchrotron emission to 
discuss the integrated spectrum of the radio relic in A2256, as it is common in the literature.\\
We have formulated five different possible scenarios to explain the properties of the relic, and we will discuss here the pros and cons for each of them.
\subsection{Non-stationary DSA}    
Together, the regions G and H of the relic in A\,2256 reach a length of $\sim$ 1 Mpc. 
To date the most plausible scenario to explain such giant radio relics invoke diffusive shock acceleration (DSA) during cluster mergers. 
In this scenario, electrons are accelerated to an injection power-law spectrum at the shock location, and lose energy due to synchrotron and IC processes
advecting downstream, causing
a local steepening of the spectrum, the entity of which changes with the distance to the shock front. The continuous injection model is used to describe the 
situation in which these different regions are not spatially resolved and their contribution is mixed up.
 Assuming that the shock is continuously accelerating particles following the same power law for a time exceeding the electrons cooling time, we expect a 
 volume-integrated spectrum that is a single straight power law with spectral index $\alpha_{obs}=\alpha_{inj}$ + 0.5 (stationarity for the spectrum). 
As shown in Sect. \ref{subsec:relic_spec}, the relic in A\,2256 shows, instead, a peculiar broken power law. 
At low frequencies (between 63 and 1369 MHz), we confirm the spectral index $\alpha \sim 0.85$ previously found by \citet{2012A&A...543A..43V}.
This spectrum is too flat to be considered the stationary spectrum as it would imply an injection spectral index $\sim 0.35$, flatter than the flattest allowed by test-particle DSA theory (see Introduction).
One solution is that, at low frequencies, we observe the injection synchrotron spectrum since a stationary spectrum is only attained if the time for energy losses is shorter than the age of the shock at all energies.
\citet{2011JKAS...44...49K} shows that the downstream integrated electron spectrum at a quasi-parallel shock is a broken power law 
which steepens from E$^{-\delta_{inj}}$ to E$^{-(\delta_{inj} + 1)}$ above E > E$_{br}(t)$, with an exponential cutoff at energies higher than $E_{eq}$. 
E$_{eq}$ is the maximum electron energy injected reached when the equilibrium between energy gains and losses is achieved and
reflects the strength of the shock. E$_{br}$ describes more properly
 the electron aging and can be used to estimate the shock age.
As a consequence the volume-integrated synchrotron spectrum has a spectral break and an high-frequency cut-off described by: 
\[
  S(\nu) \propto
  \begin{cases}
   \nu ^{- \alpha_{inj}}         & \text{at  } \nu < \nu_{br} (t) \\
   \nu ^{- (\alpha_{inj}+0.5)}   & \text{at  } \nu_{br} (t) < \nu < \nu_{eq}\\
   exp(-\nu/\nu_{eq})            & \text{at  } \nu > \nu_{eq}
  \end{cases}
\]
If we assume that the shock started to accelerate particles recently, we would be able to catch the break frequency 
and the transition from $\alpha_{inj}$ to $\alpha_{obs}$ still in the observable frequency range (non-stationarity for the spectrum), 
as it happens for young radio galaxies \citep[e.g.,][]{1999A&A...345..769M}.
Assuming a $\nu_{br} \sim$ 1.4 GHz (in correspondence of which we observe the moderate steepening) would imply a relic age of $t_{rad}(\gamma_{e,1.4GHz}) \sim 0.04 - 0.05$ Gyr.
An injection spectral index $\sim$ 0.85 would be also in agreement with what reported by \citet{2011MmSAI..82..547C}. They observe a spectral index of 0.85 at the outer edge of the relic,
steepening moving across the source towards the inner edge. This is in agreement with a scenario where the relic is produced by a shock 
that is moving outwardly and is now located at the outer edge of the relic where it is accelerating particle according to an injection spectrum 
with $\alpha_{inj}$=0.85. However, \citet{2014arXiv1408.5931O} recently found that the simple gradient from north to south in spectral index does not dominate
the relic structure when observed in more detail.
Moreover, the spectral index of the integrated synchrotron emission above the break frequency should be in this case $\alpha_{obs} \sim 1.35$, 
while we observe a flatter spectrum with slope $\sim$1. However, the spectral curvature of the continuous injection model is gradual and 
might occur over a broader range of frequencies than sampled here. 
This would place the break frequencies at frequencies even higher than 1.4 GHz, 
further reducing the radiative lifetime of the relic. This would make the radio relic in A2256 a very young radio relic.\\
One disadvantage of integrated spectra is that the spectra of different source components are mixed up. 
The radiative ages derived from these spectra do not necessarily represent
the source ages, but rather the radiative ages of the dominant source components. For radio galaxies, for example, only when the lobes (which have accumulated the 
electrons produced over the source lifetime) dominate the source spectrum, the radiative age derived from the $\nu_{br}$ is likely representative of the age of the entire 
source. If, instead, the spectrum is dominated by the jets or hot spots, the radiative age likely represents the permanence time of the electrons in that component 
and is expected to be -perhaps much- less than the source age \citep{1999A&A...345..769M}. Something similar might hold for the radio relic in A2256 if its synchrotron
emission is dominated by the shock region. This might happen because the magnetic field is expected to be stronger at the shock location due to shock compression 
\citep{2000MNRAS.314...65L}. The continuous injection model assumes, instead, a constant magnetic field over all the source, and might not be optimal to describe the integrated spectra
of radio relics. In general, if the latter are dominated by the emission from the regions closer to the shock front, they are biased to flatter values. As a consequence,
the shock Mach numbers derived from radio observable through eqs. \ref{eq:alfa_inj} - \ref{eq:delta_inj} are biased to higher values. 
This would also account for the discrepancies between the values 
of the Mach number derived from radio observations vs values derived from X-ray observations claimed in some case 
\citep[e.g., in the cluster ZwCl 2341.1+0000][]{2014MNRAS.443.2463O}.\\
The observed radio spectrum of the relic in A\,2256 may be reconciled with shock acceleration even invoking a more complex 
situation in which the electrons emitting in the relic region belong to different populations with different acceleration history .
This might be due to a modification of the shock properties
while propagating in the ICM (non-stationarity for the shock) and a consequent modification of the electron injection power law.
Indeed, shocks developing in the core of a merging event will typically strengthen moderately as they propagate into lower density and low-temperature
regions outside the cluster cores \citep{2014IJMPD..2330007B}. This results in a flattening of the injection power law as the shock propagates.\\ 
A combination of the two effects might take place in the A2256 relic, with consequences on the integrated spectrum that are difficult to quantify.
They could be responsible for the flat behavior of its integrated spectrum.

\subsection{CRe modified DSA}
The relic spectrum can be fitted also with a single power law between 63 MHz and 10450 MHz with a spectral index $\alpha_{63}^{10450}=0.92\pm0.02$.
In this case the relic in A2256 would be the one with the flattest integrated spectrum known so far, according to the collection in 
Table 4 in \citet{2012A&ARv..20...54F}.
The spectral index would be even flatter if the SZ decrement approaches our upper limits estimate.
It would be quite unlikely that we are observing the injection spectrum because this would imply that $\nu_{br} > 10$ GHz that is equivalent to say that
the relic brightened more recently than 0.016-0.019 Gyr. 
So, if what we observe is the stationary spectrum, it would imply an injection spectral index $\sim 0.42$, flatter than the flattest allowed by test particle DSA. 
Such a flat radio spectrum should be produced by a relatively strong shock, even though predictions on the Mach number are not possible inside the test-particle 
approximation of DSA. 
However, we observe a low temperature ratio of 1.7. Although projection effects might have 
reduced this value, we speculatively consider unlikely that they could have hidden a strong shock as the one required. However, the X-ray data examined in this paper 
do not allow us to firmly rule out the presence of a shock front with Mach number typical for relics. Accurate profiles of the ICM surface brightness and temperature from X-ray observations, as well as pressure profiles from deep pointed SZ observations are needed
to finally establish the presence of a shock front in the location of the radio relic in A\,2256.\\
A solution is to take into account the shock modification induced by the dynamical reaction of the accelerated particles. 
The condition on the flattest possible injection index is due to the test-particle approach to DSA.
In CR modified shocks, the compression ratio can be higher and the injection synchrotron spectra can be flatter than 0.5 \citep[e.g.,][]{2010MNRAS.402.2807B}. 
\subsection{DSA of pre-accelerated CRe}
\citet{2011ApJ...734...18K} and \citet{2012ApJ...756...97K} argue that the existence of pre-accelerated CRs electrons might alleviate the problem of inefficient injection at weak shocks.
They have suggested that the hot ICM  first goes through a series of accretion shocks of high Mach numbers before getting subjected to weaker cluster merger shock, 
and that hence the ICM should contain some pre-existing CR population. Moreover, pre-existing non thermal particles might have been produced via turbulent re-acceleration
\citep{2007MNRAS.378..245B, 2011MNRAS.412..817B} or through $p-p$ collisions of CR protons with thermal protons of the ICM \citep{2001ApJ...562..233M}.
This pre-existing CR population may contain many different electron populations and may be described by a cumulative power law f$\propto$E$^{-\delta_{pre}}$. 
They show that if the spectrum of the pre-existing population is steeper than the spectrum that could be produced by the shock ($\delta_{pre}>\delta_{inj}$), 
then the re-accelerated CR spectrum gets flattened to E$^{-\delta_{inj}}$ by DSA. In the opposite case ($\delta_{pre}<\delta_{inj}$),
the re-accelerated CR spectrum is simply amplified by the factor of $\delta_{inj}/(\delta_{inj} -\delta_{pre})$ and retains the same slope as the slope of pre-existing CRs \citep{2011ApJ...734...18K,2012ApJ...756...97K}.
Applying this to the radio relic in A\,2256, we have to assume that the spectrum we observe has the same spectral slope as the one of the pre-existing population. In this case the shock we observe would be responsible only for the amplification of the particle spectrum.
However, it is difficult to explain how the radio spectrum has kept such flat slope over a wide range of frequencies in spite of electron energy losses.
\subsection{Adiabatic compression}
The relic in A\,2256 appears to be very filamentary. 
\citet{2002MNRAS.331.1011E} presented detailed three-dimensional simulations of the passage of radio plasma cocoons filled with turbulent magnetic field through shock waves.
They showed that, on contact with the shock wave, the radio cocoons are first compressed and finally torn into filamentary structures.
Moreover, \citet{2012A&A...543A..43V} proposed a viewing angle of about 30$^{\circ}$ from edge-on and a a true distance of $\sim$0.5 Mpc from the cluster center for the radio relic. 
Both these two characteristics favor of a scenario where the relic is the result of re-energization of fossil plasma by adiabatic compression. 
On the other hand, the large size of the relic and the flat spectrum at high frequency make this scenario unlikely.
\subsection{Independent relics}
Finally, regions G and H might be another example of two independent relics, for example two radio Phoenices produced by the same shock or by different shocks front.
We have checked for different properties of the two regions calculating their individual radio integrated spectra.
Although the integrated spectrum of region G appears slightly steeper than that of region H, the two are consistently similar within the error bars and too flat to be reconciled with the observed shock. 
The absence of a pronounced steepening in the individual spectra also rules out the double phoenics origin.\\

The above discussion leads to the conclusion that the radio relic in A2256 is a very special and complex case, which requires special conditions to be 
interpreted in the framework of DSA.\\ 
We presented in the introduction the radio relic in CIZA J2242.8+5301 as a textbook example of giant radio relic where the spectral predictions of the test-particle DSA
model are nicely satisfied. Actually, it has been argued that this holds only at frequencies below 2.3 GHz, where the integrated spectrum and the spatial distribution
of spectral indeces match the predictions from the test-particle DSA continuous injection model \citep{2013A&A...555A.110S}. However, recent observations of 
the relic at 16 GHz \citep{2014arXiv1403.4255S}, shows a high-frequency steepening, with the flux density at this frequency lying 12$\sigma$ below the extrapolation of the low-frequency single power-law 
spectrum. They argue that this discrepancy can be due to different factors, e.g., an injection spectrum that is not a power law, a gradient of density and/or temperature
across the source, a non-homogeneity of the magnetic field. 
We suggest that another explanation could be that the observed steepening is the expected cut-off above $E_{eq}$. Due to the lack of flux density measurements 
between 2.3 and 16 GHz, the gradualness of the steepening can not be established and
a firm comparison with the model's expectation is not easy. Another possibility is that the low flux density is due to SZ decrement, that can be important over
the large area covered by the radio relic in CIZA J2242.8+5301.
Nevertheless, \citet{2014arXiv1403.4255S} conclude that at the moment no relic formation mechanisms can readily explain the 
high-frequency steepening and new theoretical models must be developed.\\ 
We conclude that the cases of A2256 and CIZA J2242.8+5301 show that the simple model with assumptions of continuous stationary injection, constant magnetic field 
across the relic regions and stationarity of the shock properties
are insufficient when the radio spectra are known over a broader range of frequencies. The complexity of current models invoking shock acceleration must, therefore,
be increased.
However, alternative models, e.g., magnetic reconnection, might be worth to be considered in some case, as suggested by \citet{2014arXiv1408.5931O}.

\section{Conclusions}\label{sec:conclu}
       We presented new high-frequency observations of A\,2256 performed at 2273 MHz with the WSRT and at 2640 and 4850 MHz with the Effelsberg-100m telescope.
      The high resolution of the WSRT observations highlighted that the relic can be divided in two regions of enhanced surface brightness
      (regions G and H) connected by a filamentary bridge of lower brightness emission.\\
      Combining our new observations with images available at other frequencies, we constrained the radio integrated spectrum of the radio relic in 
      A\,2256 over the widest sampled frequency range collected so far for this kind of objects (63 -10450 MHz). 
      We find that the relic keeps an unusual
      flat behavior up to high frequencies. We find that, although the relic spectrum between 
      63 and 10450 MHz is not inconsistent with a single power law with $\alpha(relic)_{63}^{10450}= 0.92\pm 0.02$, a 
      separate fit of the spectra between 63 and 1369 MHz and between 1369 and 10450 MHz shows that these two frequency ranges are 
      best represented by two different power laws, with 
      $\alpha(relic)_{63}^{1369}= 0.85\pm 0.01$ and $\alpha(relic)_{1369}^{10450}= 1.00\pm 0.02$. 
      This broken power law would require special conditions to be explained in terms of test-particle DSA, e.g., non-stationarity of 
      the spectrum and/or non-stationarity of the shock. On the other hand, the single power law would make of this 
      relic the one with the flattest integrated spectrum known so far, even flatter than what allowed in the test-particle approach to DSA.\\
      We have investigated the possibility that the G and H regions have different properties in the frequency range 351-10450 MHz.
      We find that the singular spectra of regions G and H show a similar flat trend respect the total relic with $\alpha(G)_{351}^{10450}=0.97\pm0.04$ and 
      $\alpha(H)_{351}^{10450}=0.92 \pm 0.02$.\\
      We used Suzaku and XMM-Newton X-ray observations to measure the ICM temperature in the regions across the radio relic emission.
      We find a temperature ratio $T_2/T_1 \sim 1.7$ across the G region of the radio relic.
      Although projection effects might have reduced this value, we consider unlikely that they could have hidden a strong shock as the one required
      by the flat behavior of the integrated spectrum of the entire relic, as well as of the regions G and H separately. However, the X-ray data examined in this paper 
do not allow us to firmly rule out the presence of a shock front with Mach number typical for relics.
      No temperature jumps are observed across the H region of the relic. This might be due to the projection in this area of a colder subcluster approaching the main cluster 
      from the west.\\
      The absence of a pronounced steepening in the integrated radio spectrum of the relic as well as of the regions G and H separately, rules out a phoenices
      origin for the relic/s.\\
      We conclude that the case of A2256 show that the simple shock model with assumptions of continuous stationary injection, constant magnetic field 
across the relic regions and stationarity of the shock properties
are insufficient when the radio spectra are known over a broader range of frequencies. The complexity of current models invoking shock acceleration must, therefore,
be increased.

\begin{acknowledgements}

      We thank the anonymous referee for valuable suggestions , which improved the manuscript.
      M.T., A.B. and M.B. acknowledge the financial support by the German
      \emph{Deut\-sche For\-schungs\-ge\-mein\-schaft, DFG\/} project FOR 1254.
      SRON is supported financially by NWO, the Netherlands Organization for
      Scientific Research.
      L.L. acknowledges support from the DFG through the research grant RE 1462/6 and LO 2009/1-1. 
      Basic research in radio astronomy at the Naval Research Laboratory is supported by 6.1 Base funding.\\
      The authors thank M. Brentjens for providing the radio image at 351 MHz and R. van Weeren for providing information on the flux density at 63 MHz.
      M.T. thanks F. Vazza for helpful discussions on CRs acceleration mechanism and K. Basu for helping in the evaluation of the SZ effect.\\
      We thank the P.I. of the Suzaku observations K. Hayashida.
      We also thank M. Kawaharada for providing unpublished Suzaku offset data.\\
      
      Partly based on observations with the 100-m telescope of the MPIfR (Max-Planck-Institut f\"ur Radioastronomie) at Effelsberg.
      Partly based on observations with the WSRT. The Westerbork Synthesis Radio Telescope is operated by
      ASTRON (Netherlands Foundation for Research in Astronomy) with support
      from the Netherlands Foundation for Scientific Research (NWO).\\

\end{acknowledgements}
\bibliographystyle{aa} 
\bibliography{biblio.bib} 
\begin{appendix}
\section{Derivation of the flux densities of discrete sources}\label{app:sources}
Here we report the details of the derivation of the discrete sources flux densities.
All the values are listed in Table \ref{table:pointsources}. The sources position in brackets are expressed in the J2000 coordinate system.\\

\textbf{{\itshape Source C$_{tail}$} (17 03 28, +78 39 57 )}\\
  The peak of the emission of source C coincides with an optical galaxy with magnitude 15.3 \citep{1979A&A....80..201B}.
 The head-tail morphology for source C was first suggested by \citet{1976A&A....52..107B}. 
 We modeled the source distinguishing between the head (long 75$^{\prime\prime}$.6) and the tail as we are interested 
 only on the contribution from the latter.
 At the highest frequency (10450 MHz) the head of the source is still clearly visible, while it's difficult to establish if the flux in the tail region is from the 
 tail or from the underlying relic. The measured fluxes are $S(head)_{10450}=2.1 \pm 0.6$ mJy and $S(tail)_{10450}=1.9 \pm 0.7$ mJy and consequently 
 $S(all)_{10450}=4.0 \pm 0.0$; the last two can be considered as upper limits.
 Combining this values with the fluxes we measured at 2273 MHz ($S(all)_{2273}=33.7\pm 1.8$ mJy; $S(head)_{2273}=17.3\pm 0.9$ mJy ;
 $S(tail)_{2273}=16.4 \pm 0.9$ mJy), we obtain 
 spectral indeces $\alpha(all)_{2273}^{10450}=1.40 \pm 0.14$, $\alpha(head)_{2273}^{10450}=1.40\pm 0.19$ and $\alpha(tail)_{2273}^{10450}=1.42 \pm 0.25$. 
 We used these values of the spectral indeces 
 to extrapolate the fluxes of the components at 2640 and 4850 MHz. \citet{2009ApJ...694..992L} report a measured flux for the
 source C (referred to in the paper as 1706+787) of 5.06$\pm$0.60 mJy at 4.9 GHz, classifying the source as a point source. 
 Due to the wrong classification, we believe that their observations
 resolve out the tail of the source and the reported flux refers to the source head only. 
 Indeed, their reported flux is in agreement with the flux that we extrapolated for the head of the source ($S(head)_{4850}=6.0 \pm 0.9$ mJy).\\
 \citet{1994ApJ...436..654R} report a flux for the entire source at 327 MHz of 246 $\pm$ 20 mJy; combining this with our measurements at 1369 MHz ($S(all)_{1369}=56.5\pm$ 3.4)
 we find that the spectral index 
 for the entire source keeps the same slope as in the range 1369-2273 MHz, $\alpha(all)_{327}^{1369}=1.03 \pm 0.07$. 
 To separate the emission between the head and the tail we used the spectral
 index profile along the tail of the source C between 327 and 1447 MHz published by \citet[][Fig.7 Right side]{1994ApJ...436..654R}. From the plot we deduced that the averaged spectral index of the head 
 (selecting the range 0 - $\sim$75 $^{\prime\prime}$ in the x axes) is $\alpha(head)_{327}^{1369}=0.65\pm0.10$. From this we got a flux at 327 MHz for the head of 63.4 mJy and consequently a flux for the
 tail of $S(tail)_{327}=246.0 - 63.4 = 182.6$ and a spectral index $\alpha(tail)_{327}^{1369}=1.23$. We used this spectral index to calculate the fluxes at 351 MHz.\\ 
 At 153 MHz \citet{2012A&A...543A..43V} report a flux for the enire source C of $S(all)_{153}=480 \pm 50$ mJy. The resulting spectral index is
 $\alpha(all)_{153}^{327}=0.88\pm0.17$. From Figure 5.4 (Right panel) of \citet{2009PhDT.........1I} we deduce a value for the spectral index between 153 and 325 MHz 
 for the head of the source of 
 $\alpha(head)_{153}^{325}=0.50 \pm 0.10$. We used this value to calculate the flux of the head $S(head)_{153}=92.7 \pm 10.7$ mJy. Consequently the tail has $S(tail)_{153}= 387.3 \pm 51.1$ mJy and
 $\alpha(tail)_{153}^{325}=1.00$. We estrapolated the fluxes to 63 keeping costant the spectral indeces of the entire source and of the head. For the tail we obtain $S(tail)_{63}=903.5 \pm$ 257.0 mJy.\\
 
 \textbf{{\itshape Source K} (17 02 18.4, +78 46 03.30)}\\
 Source K was first identified by \citet{1979A&A....80..201B} and can be associated with a star-forming galaxy \citep{2003AJ....125.2393M}. \\
 \citet{1994ApJ...436..654R} report a flux for the source of 7.0$\pm$1.0 mJy at 327 MHz and a flux of 3.4$\pm$0.3 mJy at 1446 MHz. The latter is in agreement
 within the error bars with what we measured at 1369 MHz (3.3 $\pm$ 0.2 mJy).
 Combining these values with what we find at 2273 MHz (1.9$\pm$0.2 mJy) we obtain a spectral index $\alpha_{327}^{1446}=0.52 \pm 0.07$
 with a steepening at high frequency $\alpha_{1369}^{2273}=1.09 \pm 0.02$. We extrapolated the fluxes at 63 MHz and 351 MHz using $\alpha_{327}^{1446}$ and 
 the fluxes at 2640 and 4850 MHz
 using $\alpha_{1446}^{2273}$. The contribution of the source to the total flux at 10450 MHz is negligible ($\ll 1$ mJy). \\
 
\textbf{{\itshape Source J} (17 01 12, +78 43 27)}\\
 Source J was first identified by \citet{1979A&A....80..201B}. Measuring the fluxes at 1369 and 2273 MHz, we found that the source has an inverted-spectrum,
 in agreement with what reported by \citet{1994ApJ...436..654R}. The source is indeed not visible at the 327 MHz (\citet{1994ApJ...436..654R} report an upper limit of 1 mJy), 
 but become visible at 1369 MHz and its flux increase at 2273 MHz. At 10450 MHz the source is again at a level $<$1 mJy.
 Fitting our measurements with the flux reported by \citet{1994ApJ...436..654R} at 327 and 1447 MHz, we find a spectral index $\alpha_{327}^{2273}=-0.25 \pm 0.15$ and $\alpha_{2273}^{10450} \geq 0.35$. 
 We used this value to extrapolate the fluxes at 2640, 4850 and 10450 MHz.\\
 
\textbf{{\itshape Source I} (17 00 52.68, +78 41 23)}\\
 Source I is a head tail source first identified by \citet{1979A&A....80..201B}. The flux we measured at 1369 MHz ($9.3 \pm 0.8$) is in agreement with what
 reported in the literature at close frequencies ($S_{1447}=8.2 \pm 1.8$ mJy from \citet{1994ApJ...436..654R}; $S_{1400}=10$ mJy from \citet{1997ApJS..108...41O}, included in the fit). 
 At 2273 MHz we measure a flux of 6.8$\pm$0.4 mJy
 that imply a spectral index $\alpha_{1369}^{2273}=0.69 \pm 0.10$. At higher frequencies the source has been observed with the VLA by \citet{2009ApJ...694..992L} 
 (and classified as extended source). They
 report $S_{4900}=2.14 \pm 0.45$ mJy and $S_{8500}=1.25 \pm 0.36$ mJy. Fitting these values with our measurement at 2273 MHz we find a spectral index $\alpha_{2273}^{8500}=1.38 \pm 0.12$ that
 we used to extrapolate the fluxes at 2640, 4850 and 10450 MHz. At 327 MHz \citet{1994ApJ...436..654R} report a flux of 8 $\pm$1 mJy
 that would imply an inverted spectrum. However in their published map only the head is clearly visible, while the tail is resolved out.
 We treated their reported flux as a lower limit and we averaged it with the flux resulting from the extrapolation with the spectral index $\alpha_{1369}^{2273}$ found ($25.9$ mJy), 
 that represents, on the other hand, an upper limit. The resulting averaged flux is $S_{327}=17.0 \pm 8.9$ and consequently $\alpha_{327}^{1447}=0.39 \pm 0.21$ mJy. We used this value to estrapolate the fluxes at 
 63 and 351 MHz. \\
 
\textbf{{\itshape Source I2} (17 01 24, +78 41 13)}\\
  We measure, for source I2, a flux of $1.00 \pm 0.07$ mJy at 1369 MHz and $1.9 \pm 0.1$ mJy at 2273 MHz implying an inverted 
  spectrum. The source is indeed invisible in the 327 MHz VLA map published by \citet{1994ApJ...436..654R}.
  At 10450 MHz we measure an upper limit $<$1 mJy. \\
  
\textbf{{\itshape Sources K2 (17 02 30, +78 45 00), J2 (17 00 51, +78, 42, 28), I3 (17 02 02, +78 40 32), G2 (17 03 22.7, +78 46 56.1)}}\\
   For these sources only our measurements at 1369 MHz and 2273 MHz are available. We extrapolated the fluxes at the other frequencies assuming
   straight spectra. 
\end{appendix}
\end{document}